\documentclass[12pt]{article}

\usepackage{jheppub}
\usepackage{amsbsy,amsfonts,amsmath,amssymb}
\usepackage{cleveref} 
\usepackage[g]{esvect} 

\crefname{equation}{}{}
\crefname{chapter}{Chapter}{Chapters}
\crefname{section}{Section}{Sections}
\crefname{subsection}{Subsection}{Subsections}
\crefname{subsubsection}{Subsubsection}{Subsubsections}
\crefname{figure}{Figure}{Figures}
\crefname{table}{Table}{Tables}
\crefname{appendix}{Appendix}{Appendices}

\renewcommand{\arraystretch}{1.5} 

\allowdisplaybreaks

\title{Second order transport coefficients of nonconformal fluids from compactified Dp-branes}

\author[]{Chao Wu and Yanqi Wang}

\affiliation[]{School of Physics and Materials Science, Anhui University, Hefei 230601, China}

\emailAdd{chaowu86@outlook.com, wangyanqi0@gmail.com}

\abstract{All the 7 dynamical second order transport coefficients of the nonconformal fluids that correspond to Dp-branes with one or more world-volume directions compactified are derived via fluid/gravity correspondence. The conditions considered in this paper include D4-brane with 1, 2 or 3 compact directions, D3-brane with 1 or 2 compact directions, as well as D2-brane with 1 direction compactified. The derived second order transport coefficients satisfy the Haack-Yarom, Romatschke and Kleinert-Probst relations.}


\arxivnumber{2012.14699}

\begin{document}

\maketitle

\section{Introduction}

Fluid/gravity correspondence \cite{Bhattacharyya0712,Bhattacharyya0803} is an effective method of calculating the dynamical second order transport coefficients for various kinds of gravity backgrounds. After it was proposed, this framework has been used in asymptotically Anti de Sitter (AdS) black holes in various dimensions \cite{VanRaamsdonk0802,Haack0806,Bhattacharyya0809}, with background scalar field \cite{Bhattacharyya0806} and with background vector charge \cite{Erdmenger0809,Banerjee0809}.

Branes in string/M theory are the vacuum solutions of 10/11 dimensional supergravity \cite{Duff9412}. The methods used in previous studies on the transport properties of branes are the Minkowski AdS/CFT correspondence \cite{Herzog0210,Parnachev0506,Benincasa0605,Mas0703,Natsuume0712,Natsuume0807,David0901}, the stretched horizon formalism \cite{Kapusta0806,Springer0810,Springer0902} and the boundary derivative expansion method in Fefferman-Graham coordinate \cite{Kanitscheider0901}. The most remarkable observation that one can make from these studies is that the relativistic fluids dual to brane systems are nonconformal since background dilaton exists.
Following \cite{Baier0712}, reference \cite{Romatschke0906} proposes the most general formulation of the second order constitutive relations in nonconformal regime, which sets the stage for later systematic investigations on nonconformal relativistic fluids.

Based on these previous works, \cite{Wu1508,Wu1604,Wu1608,Wu1807} calculate the first and second order transport coefficients for D-branes via the fluid/gravity correspondence and have got some new results. Of those references, \cite{Wu1508,Wu1604} study the D4-brane with 1 world-volume direction compact and \cite{Wu1807} calculates all the dynamical second order transport coefficients for Dp-brane. This paper can be seen as a follow-up work of \cite{Wu1807} and we are aiming to calculate all the dynamical second order transport coefficients for Dp-brane with $q$ of its world-volume direction(s) compactified, where $1\leq q\leq p-1$. This can be seen as our first motivation. The maximum of $q$ can not be $p$ since we need to leave at least one dimension on the world-volume for the fluid to live in. We denote a Dp-brane with $q$ direction(s) compact as D(p-q)-brane, such that the cases considered here are D(4-1), D(4-2), D(4-3), D(3-1), D(3-2) and D(2-1)-brane. Actually, the D(4-1) case has been studied in \cite{Wu1508,Wu1604}, thus we will not offer the details for this case but will include its results in the final expression for the second order constitutive relation.

Our second motivation is to give a final answer to the question proposed in \cite{Mas0703} which calculates the first order transport coefficients for Dp-brane. In the conclusion of that paper, in order to make their result compatible with that of \cite{Benincasa0605}, the authors proposed an expression for the radio $\zeta/\eta$. We quote it here in our convention as
\begin{align}\label{eq: zeta over eta in Mas0703}
  \frac \zeta \eta = \frac{2(p-3)^2 + 2q(5-p)}{(p-q)(9-p)}.
\end{align}
After discovering the above can cover both their results and that of \cite{Benincasa0605}, the authors of \cite{Mas0703} say that they do not know whether this compatibility is just a coincidence or due to some reason behind the formula. Now we can answer this question: the above expression of $\zeta/\eta$ is for the compactified Dp-brane, or in our notation, the D(p-q)-brane. When setting $q=0$, it will be back to the cases in \cite{Mas0703} and when setting $p=4,~q=1$, it reproduces the result of \cite{Benincasa0605}.

The structure of the paper is as follows. In \cref{sec: dimensional reduction} we will reduce the 10 dimensional background of near-extremal black Dp-brane to a $p-q+2$ dimensional Einstein-dilaton theory. In \cref{sec: the 1st order} we will answer the question proposed in \cite{Mas0703} by offering the result of the first order constitutive relation for D(p-q)-brane. \cref{sec: preliminaries on the 2nd order} is a preparation for the second order and then in \cref{sec: solve the 2nd order perturbations} we will solve the differential equations for all the second order perturbations. The final result of second order stress energy tensor and transport coefficients will be given in \cref{sec: constitutive relations of the 2nd order}. \cref{sec: summary} is a summary about this paper.

\section{The reduction from 10 to $\boldsymbol{p-q+2}$ dimensions} \label{sec: dimensional reduction}

This section shows how to reduce the 10 dimensional background of Dp-brane to a $p-q+2$ dimensional spacetime. Since this work is subsequent to \cite{Wu1807}, the conventions are the same as that reference. We will then be brief in offering the details of the definitions which are the same as before and will also stress the differences.

The action for Dp-brane in 10 dimension contains the bulk term, the Gibbons-Hawking surface term and the counter term. It reads as
\begin{align}
  S =&\; \frac1{2\kappa_{10}^2} \int d^{10}x \sqrt{-G} \left[ \mathcal R - \frac12 (\nabla_{\hat M}\phi)^2 - \frac{g_s^2}{2(8-p)!} e^{\frac{p-3}2\phi} F_{8-p}^2 \right] \cr
  & - \frac{1}{\kappa_{10}^2} \int d^9x \sqrt{-H} \mathcal K + \frac{1}{\kappa_{10}^2} \int d^9x \sqrt{-H} \frac{9-p}{2L} e^{\frac{3-p}{4(7-p)}\phi}.
\end{align}
The equations of motion (EOM) derived from this action admit the following background solution:
\begin{align}
   ds^2 &= \left[ \left( \frac r{L_p} \right)^\frac{(7-p)^2}8 \Big( -f(r)dt^2 + \delta_{ij}dx^idx^j \Big) + \left( \frac{L_p}r \right)^\frac{(p+1)(7-p)}8 \frac{dr^2}{f(r)} \right] \cr
  & + \left( \frac r{L_p} \right)^\frac{(7-p)^2}8 \delta_{mn}dy^mdy^n + \left( \frac{L_p}r \right)^\frac{(p+1)(7-p)}8 r^2 d\Omega_{8-p}^2, \label{eq: 10d Dp metric} \\
   e^\phi &= \left( \frac r{L_p} \right)^\frac{(p-3)(7-p)}4, \qquad \qquad F_{\theta_1 \cdots \theta_{8-p}} = g_s^{-1} Q_p \sqrt{\gamma_{8-p}},
\end{align}
where $f(r) = 1 - \frac{r_H^{7-p}}{r^{7-p}}$ and $Q_p = (7-p)L_p^{7-p}$. Here $L_p^{7-p} = \frac{(2\pi l_s)^{7-p} g_s N}{(7-p) \Omega_{8-p}}$, which relates the charge of the Dp-brane and will become the characteristic length of the reduced spacetime.

The 10 dimensional line element \cref{eq: 10d Dp metric} has three parts and represents a spacetime with structure of $\mathcal M_{p-q+2}\times \mathbf T^q \times \mathbf S^{8-p}$, of which the coordinate is $x^{\hat M}=(x^M, y^m, \theta^a)$. Therein, $x^M=(x^\mu, r)$ are the coordinates of $\mathcal M_{p-q+2}$ with $\mu=0,1,\cdots,p-q$; $y^m$ are the coordinates of the compactified directions on D-branes that forms a $q$-dimensional torus $\mathbf T^q$ with $m=1,\cdots,q$; and $\theta^a$ are the coordinates of $8-p$ dimensional sphere $\mathbf S^{8-p}$ with $a=1,\cdots,8-p$.

In the following we will implement the dimensional reduction on the compact dimensions of the gravity side $\mathbf T^q$ and $\mathbf S^{8-p}$ by integrating the coordinates $y^m$ and $\theta^a$, respectively. The dimensional reduction on these two compact subparts of the bulk metric has different significance. The reduction on $\mathbf S^{8-p}$ is just to reconcile the Dp-brane metric with the framework of fluid/gravity correspondence. The reduction on $\mathbf T^q$ can modify the metric of the reduced bulk gravity, hence causing changes on the results of the second order transport coefficients.

We use the ansatz
\begin{align}\label{eq: 10D reduction ansatz}
  ds^2 = e^{2\alpha_1 A} g_{MN} dx^M dx^N + e^{2\alpha_2 A}\left( e^{2\beta_1 B} \delta_{mn}dy^m dy^n + e^{2\beta_2 B} L_p^2 d\Omega_{8-p}^2 \right)
\end{align}
to perform the dimensional reduction. If we denote the metric for the above ansatz as $G_{\hat M \hat N} = \{ e^{2\alpha_1 A} g_{MN},~e^{2\alpha_2 A + 2\beta_1 B} \delta_{mn},~e^{2\alpha_2 A + 2\beta_2 B} L_p^2 \gamma_{ab} \}$ then the 9D induced boundary metric is $H_{\hat M\hat N} = G_{\hat M\hat N} - \mathbf n_{\hat M} \mathbf n_{\hat N}$ and the corresponding line element is
\begin{align}\label{eq: 9D reduction ansatz}
  ds^2 = e^{2\alpha_1 A} h_{MN} dx^M dx^N + e^{2\alpha_2 A}\left( e^{2\beta_1 B} \delta_{mn}dy^m dy^n + e^{2\beta_2 B} L_p^2 d\Omega_{8-p}^2 \right),
\end{align}
where $h_{MN}$ is the boundary metric of $\mathcal M_{p-q+2}$. The unit norm on the 9D boundary is
\begin{align}
  \mathbf n_{\hat M} = \frac{\nabla_{\hat M}r}{\sqrt{ G^{\hat N\hat P} \nabla_{\hat N}r \nabla_{\hat P}r }} = (\mathbf n_M, \mathbf n_m, \mathbf n_a) = \left( e^{\alpha_1 A} n_M, 0, 0 \right),
\end{align}
with $n_M$ the unit norm on the boundary of $\mathcal M_{p-q+2}$.

As said in \cite{Wu1508}, the requirement that the reduced action is also in Einstein frame gives two relations among those 4 parameters in \cref{eq: 10D reduction ansatz}. So one may set them as
\begin{align}
  \alpha_1 = - \frac{8-p+q}{p-q}, \qquad \alpha_2 = 1, \qquad \beta_1 = - \frac{8-p}{q}, \qquad \beta_2 = 1.
\end{align}
The reduced action for the $p-q+2$ dimensional theory turns out to be
\begin{align}\label{eq: (p-q+2) dimensional reduced action}
  S =&\; \frac1{2\kappa_{p-q+2}^2} \int d^{p-q+2}x \sqrt{-g} \left[ R - \frac12 (\partial \phi)^2 - \frac{8(8-p+q)}{p-q} (\partial A)^2 \right.  \cr
  &\left. - \frac{(8-p)(8-p+q)}{q} (\partial B)^2 + V(\phi,A,B) \right] - \frac1{\kappa_{p-q+2}^2} \int d^{p-q+1}x \sqrt{-h} K \cr
  & + \frac1{\kappa_{p-q+2}^2} \int d^{p-q+1}x \sqrt{-h} \frac{9-p}{2L_p} \exp\left[ - \frac{8-p+q}{p-q} A + \frac{3-p}{4(7-p)} \phi \right], \\
  V(\phi,A,B) =&\; \frac{(7-p)(8-p)}{L_p^2} \exp\left( - \frac{16}{p-q}A - 2B \right) \cr
   & - \frac{Q_p^2}{2L_p^{2(8-p)}} \exp\left[ \frac{p-3}{2} \phi - \frac{2(p-q)(7-p) + 16}{p-q}A - 2(8-p) B \right].
\end{align}
In the above,
\begin{align}
  \frac1{2\kappa_{p-q+2}^2} = \frac{L_p ^{8-p} \Omega_{8-p} V_q}{2\kappa_{10}^2},
\end{align}
where $V_q = \int d^q y = \beta_1 \cdots \beta_q $ is the volume of the compact directions on the Dp-brane world-volume. $2\kappa_{p-q+2}^2$ is the surface gravity of the $p-q+2$ dimensional reduced gravity theory and it relates with that of Dp-brane \cite{Wu1807} by
\begin{align}
  \frac1{2\kappa_{p-q+2}^2} = \frac{V_q}{2\kappa_{p+2}^2}.
\end{align}

The reduced background fields that solve the reduced action (\ref{eq: (p-q+2) dimensional reduced action}) are
\begin{align}
  ds^2 &= \left( \frac r{L_p} \right)^{\frac{9-p}{p-q}} \left( -f(r) dt^2 + d \vv x^2 \right) + \left( \frac{r}{L_p} \right)^\frac{(p^2-8p+9) + q(7-p)}{p-q} \frac{dr^2}{f(r)},  \label{eq: (p-q+2) dimensional reduced metric} \\
  e^{\phi} &= \left( \frac r{L_p} \right)^\frac{(p-3)(7-p)}{4}, \label{eq: dilaton of reduced background} \\
  e^{A} &= \left( \frac r{L_p} \right)^{\frac{(p-3)^2}{16} + \frac{q(5-p)}{2(8-p+q)}}, \quad e^B = \left( \frac r{L_p} \right)^{- \frac{q(5-p)}{2(8-p+q)}}. \label{eq: A and B of reduced background}
\end{align}
For the cases of $p=3$, the scalar fields $A,~B$ still not vanish. This suggests that the D3-brane will correspond to nonconformal relativistic fluids after compactification. To put it more generalized, dimensional reduction can make the dual fluid of a conformal background nonconformal. This tells us that we can perform a dimensional reduction on AdS black hole backgrounds and calculate the second order transport coefficients of the dual fluids, which may give a direct check on the proposal in \cite{Kanitscheider0901}.

The reduced theory is actually a $p-q+2$ dimensional Einstein gravity coupled with one background scalar---for which one can choose the dilaton. This is because the three scalars in \cref{eq: dilaton of reduced background,eq: A and B of reduced background} are not independent as can be seen from the following:
\begin{align}\label{eq: A and B proportional to phi}
  A &= \left[ \frac{p-3}{4(7-p)} + \frac{2q(5-p)}{(8-p+q)(p-3)(7-p)} \right] \phi, \cr
  B &= \frac{- 2q(5-p)}{(8-p+q)(p-3)(7-p)} \phi.
\end{align}
The above are not appropriate for $p=3$ because of the terms $p-3$ in the denominator. But from \cref{eq: A and B of reduced background} one can figure out that $A = - B$ when $p=3$ and dilaton vanishes. Anyway, there is always just one independent scalar field in the background of the $p-q+2$ dimensional reduced gravity theory.

In Eddington-Finkelstein coordinate, the boosted form of \cref{eq: (p-q+2) dimensional reduced metric} is
\begin{align}
  ds^2 &= - r^\frac{9-p}{p-q} f(r) u_\mu u_\nu dx^\mu dx^\nu + r^\frac{9-p}{p-q} P_{\mu\nu} dx^\mu dx^\nu - 2 r^\frac{(p-3)(p-6) + q(7-p)}{2(p-q)} u_\mu dx^\mu dr, \cr
  u^\mu &= \frac{(1, \vv \beta)}{\sqrt{1-\vv \beta^2}}, \qquad P_{\mu\nu} = \eta_{\mu\nu} + u_\mu u_\nu
\end{align}
where now $x^0 = v$ is the in-going Eddington time. The above represents ideal relativistic fluid living on the boundary of $\mathcal M_{p-q+2}$. The fluid/gravity correspondence \cite{Bhattacharyya0712,Bhattacharyya0803} states that to mimic the real fluid flow, one should first promote the boost parameter $u^\mu$ and $r_H$ to be $x^\mu$ dependent, then add the (also $x^\mu$ dependent) metric perturbations, and finally solve them from EOM in the order of partial derivatives of $x^\mu$ in some local patch on the boundary. The global solution of the metric that is dual to arbitrary boundary flow can then be got by stitching together all local solutions. The boundary coordinate dependent global metric ansatz can be set as
\begin{align}\label{eq: x dependent global metric}
  ds^2 =& -r^\frac{9-p}{p-q} [ f(r_H(x),r) - k(r_H(x), u^\alpha(x), r) ] u_\mu(x) u_\nu(x) dx^\mu dx^\nu \cr
  &+ 2 r^\frac{9-p}{p-q} P_\mu^\rho(u^\alpha(x)) w_\rho(u^\alpha(x), r) u_\nu(x) dx^\mu dx^\nu  \cr
  & + r^\frac{9-p}{p-q} [ P_{\mu\nu}(u^\alpha(x)) + \alpha _{\mu\nu}(r_H(x), u^\alpha(x), r) + h(r_H(x), u^\alpha(x), r) P_{\mu\nu}(u^\sigma(x)) ] dx^\mu dx^\nu \cr
  & - 2 r^\frac{(p-3)(p-6) + q(7-p)}{2(p-q)} [ 1 + j(r_H(x), u^\alpha(x), r) ] u_\mu(x) dx^\mu dr.
\end{align}
This metric contains not only the boosted black brane metric but also the artificially set perturbations, which are the tensor perturbation $\alpha _{\mu\nu}(r_H(x), u^\alpha(x), r)$, the vector perturbation $w_\rho(u^\alpha(x), r)$ and the scalar perturbations $k(r_H(x), u^\alpha(x), r)$, $h(r_H(x), u^\alpha(x), r)$ and $j(r_H(x), u^\alpha(x), r)$. They depend on the boundary coordinates $x^\mu$ through the proper velocity $u^\mu$ and the horizon parameter $r_H$. All the perturbations will be solved tubewisely \cite{Bhattacharyya0712} in the bulk of the reduced $p-q+2$ dimensional nonconformal spacetime by choosing some special point in $x^\mu$ directions---which is just $x^\mu=0$, following \cite{Bhattacharyya0712}.

The tubewise solving procedure of \cref{eq: x dependent global metric} is preformed by expanding it in terms of derivatives of boundary coordinates of $u^\mu$ and $r_H$. The EOM derived from \cref{eq: (p-q+2) dimensional reduced action} that \cref{eq: x dependent global metric} satisfies order by order reads
\begin{align}
   E_{MN} &- T_{MN} = 0, \label{eq: EOM Einstein eq} \\
   \nabla^2\phi &- \frac{(p-3)(7-p)^2}{4L_p^2} \exp\left[ \frac{p-3}2 \phi - \frac{2(p-q)(7-p) + 16}{p-q} A - 2(8-p) B \right] = 0,
   \label{eq: EOM dilaton} \\
   \nabla^2 A &- \frac{(7-p)(8-p)}{(8-p+q) L_p^2} e^{- \frac{16}{p-q}A - 2B} + \frac{(7-p)^2 [(p-q)(7-p) + 8]}{16(8-p+q) L_p^2} \cr
  & \times \exp\left[ \frac{p-3}2 \phi - \frac{2(p-q)(7-p) + 16}{p-q} A - 2(8-p) B \right] = 0, \label{eq: EOM A} \\
   \nabla^2 B &- \frac{q(7-p)}{(8-p+q) L_p^2} e^{- \frac{16}{p-q}A - 2B} + \frac{q(7-p)^2}{2(8-p+q) L_p^2} \cr
   & \times \exp\left[ \frac{p-3}2 \phi - \frac{2(p-q)(7-p) + 16}{p-q} A - 2(8-p) B \right] = 0, \label{eq: EOM B}
\end{align}
where $E_{MN}$ is the Einstein tensor and the energy momentum tensor on the gravity side is defined as
\begin{align}
  T_{MN} =&\; \frac12 \left( \partial_M \phi \partial_N \phi - \frac12 g_{MN} (\partial \phi)^2 \right) + \frac{8(8-p+q)}{p-q} \left(\partial_M A \partial_N A - \frac12 g_{MN} (\partial A)^2 \right)  \cr
  & + \frac{(8-p)(8-p+q)}{q} \left(\partial_M B \partial_N B - \frac12 g_{MN} (\partial B)^2 \right) + \frac12 g_{MN} V.
\end{align}
The EOM of $\phi,~A$ and $B$ are actually not independent from each other. This is because both $A$ and $B$ are proportional to $\phi$, as can be seen from \cref{eq: A and B proportional to phi}. The consequence of this fact is that the differential equations that derived from \cref{eq: EOM A,eq: EOM B,eq: EOM dilaton} are the same, as long as we do not open the perturbations for $\phi$, $A$ and $B$. We will take advantage of this in solving the scalar perturbations for the compactified D3-brane: since there is no dilaton's EOM at hand for D3-brane, we will use the EOM of $A$ instead.

\section{The first order calculation} \label{sec: the 1st order}

\subsection{Solving the perturbations}

We need only to solve the compactified Dp-brane with $2\leq p\leq 4$. Because D5 and D6-brane do not have dual fluids, as has been pointed out in \cite{Wu1807} and references therein. D1-brane can not be compactified since it has only one spatial dimension. We need at least 1 spatial dimension for the relativistic fluids to live in and support their viscous terms.

We expand \cref{eq: x dependent global metric} to the first order of the derivative of $x^\mu$ at $x^\mu = 0$ and get
\begin{align}\label{eq: 1st order expanded metric}
  ds^2 =&\; r^\frac{9-p}{p-q} \bigg[ - \bigg( f(r) - \frac{(7-p)r_H^{6-p}}{r^{7-p}}\delta r_H - k^{(1)}(r) \bigg) dv^2 \cr
  & + 2 \big( (f-1) \delta \beta_i - w^{(1)}_i(r) \big) dvdx^i  + (\delta_{ij} + \alpha_{ij}^{(1)}(r) + h^{(1)}(r) \delta_{ij}) dx^idx^j \bigg] \cr
  & + 2 r^\frac{(p-3)(p-6) + q(7-p)}{2(p-q)} (1 + j^{(1)}(r)) dvdr - 2 r^\frac{(p-3)(p-6) + q(7-p)}{2(p-q)} \delta\beta_i dx^idr
\end{align}
All the perturbations depend only on $r$ after derivative expansion, which coincides with the fact that the fluid within the patch associated with that point is in local equilibrium. The above first order expanded metric depends on $x^\mu$ linearly through $g^{(1)}_{vv}$, $g^{(1)}_{vi}$ and $g^{(1)}_{ir}$, which contains $\delta r_H = x^\mu \partial_\mu r_H$ and $\delta\beta_i = x^\mu \partial_\mu \beta_i$. The dependence on $r$ is much more complicated than on $x^\mu$ since the perturbations will turn out to be complicate functions of $r$. The perturbations will be solved in groups of how they transform under $SO(p-q)$. All the components of the traceless tensor and vector perturbations are on the same footing because of the $SO(p-q)$ invariance.

The tensor perturbation satisfies the traceless tensor part of the Einstein equation
\begin{align}\label{eq: Einstein eq(ij)}
  E_{ij} - \frac1{p-q} \delta_{ij} \delta^{kl} E_{kl} - \left( T_{ij} - \frac1{p-q} \delta_{ij} \delta^{kl} T_{kl} \right) &= 0,
\end{align}
After we put \cref{eq: 1st order expanded metric} in, it gives
\begin{align}
  \partial_r (r^{8-p} f(r) \partial_r \alpha^{(1)}_{ij}(r)) + (9-p) r^\frac{7-p}2 \sigma_{ij} = 0 \label{eq: diff eq of 1st order alpha(ij)}
\end{align}
which does not depend on $q$. The reason is that the compactification on world-volume directions will decrease the spatial dimensions of Dp-brane. This will cause the tensor and vector perturbations losing some of their components.
But compactification will not change the differential equations of the tensor and vector perturbations since every component of them satisfies the same differential equation because of the isotropy in all spatial directions.
The existing components of the tensor and vector perturbations satisfy the same differential equations as the reduced ones.
Compactification will only affects the spatial trace of the bulk metric since taking the spatial trace is summing over all spatial dimensions. Considering $\alpha_{ij}$ is the traceless part of the metric perturbations and does not relate to the sum of spatial components, it should be no wonder that \cref{eq: diff eq of 1st order alpha(ij)} does not depend on $q$.

The SO$(p-q)$ invariance of the bulk metric allows us to set $\alpha_{ij}^{(1)} = F(r) \sigma_{ij}$, then
\begin{align}
  (r^{8-p} f F')' + (9-p) r^\frac{7-p}{2} = 0.
\end{align}
The solution of the above has been solved in \cite{Wu1807} as
\begin{align} \label{eq: Dp-brane F(r)}
  F(r) =&\; \frac1{3r_H^{1/2}} \left[ 2\sqrt3 \arctan\frac{ \sqrt{3r r_H} }{r - r_H} \right. \cr
  & \left. + \ln{ (\sqrt r + \sqrt r_H)^4 (r + \sqrt{rr_H} + r_H)^2 (r^2 + r r_H + r_H^2) \over r^6 } \right], \quad \text{(D4-brane)} \cr
  F(r) =&\; \frac1{2r_H} \left[ 2\arctan\frac{r_H}{r} + \ln\frac{(r + r_H)^2 (r^2 + r_H^2)}{r^4} \right], \quad \text{(D3-brane)} \cr
  F(r) =&\; \frac2{5r_H^{3/2}} \Bigg[ 2\sin\frac{2\pi}5 \arctan{ 2\sin\frac\pi5 \sqrt{rr_H}\over r - r_H} - 2\sin\frac\pi5
  \arctan{ 2\sin\frac{2\pi}5 \sqrt{rr_H} \over r-r_H} \cr
  & - 2\cos\frac{2\pi}5 \,{\rm artanh}\, \frac{2\cos\frac{\pi}5 \sqrt{rr_H}}{r+r_H} - 2\cos\frac\pi5 \,{\rm artanh}\,\frac{2\cos\frac{2\pi}5 \sqrt{rr_H}}{r+r_H}  \cr
  &+ \ln\frac{(\sqrt r + \sqrt{r_H})^2 (r^4 + r^3r_H + r^2r_H^2 + rr_H^3 + r_H^4)}{r^5} \Bigg]. \quad \text{(D2-brane)}
\end{align}

There are two components of the Einstein equation can be used to solve the vector perturbation: the $(0i)$ and $(ri)$ components. In the original framework, $(ri)$ component of Einstein equation
\begin{align}
  E_{ri}-T_{ri} = 0 \label{eq: Einstein eq(ri)}
\end{align}
 is used to solve $w ^{(1)}_i$ as
 \begin{align}\label{eq: 1st order w_i solution}
   w_i^{(1)}(r) = a(r) \partial _0\beta_i, \qquad a(r) = - \frac{2}{(5-p) r^\frac{5-p}2}
 \end{align}
 The linear combination of the $(0i)$ and $(ri)$ components of Einstein equation
 \begin{align}
   g^{r0} (E_{0i}-T_{0i}) + g^{rr} (E_{ri}-T_{ri}) = 0, \label{eq: Einstein eq (0i)+(ri)}
 \end{align}
 is used to give a constraint relation between the two first order vector viscous terms as
\begin{align}
   \frac{1}{r_H} \partial _ir_H = - \frac{2}{5-p} \partial _0 \beta_i, \label{eq: 1st order vector constraint}
\end{align}
Both the first order vector perturbation \cref{eq: 1st order w_i solution} and constraint equation \cref{eq: 1st order vector constraint} are not $q$-dependent. The reason for the vector perturbation has been explained below \cref{eq: diff eq of 1st order alpha(ij)}. The vector constraint equation \cref{eq: 1st order vector constraint} is a linear combination of the vector components of Einstein equation, which should also not be $q$-dependent following similar argument as what is below \cref{eq: diff eq of 1st order alpha(ij)}.

To solve the scalar perturbations, we have $(00),~(0r),~(rr)$ and $(ii)$ (with $i$ summed) components of Einstein equation, as well as the EOM of $\phi$. But there are only 3 unknown scalar functions, thus two of the five equations are redundant. The $(ii)$ (with $i$ summed) component turns out to be more complex than the others thus can be omitted. A linear combination of the $(00)$ and $(0r)$ components of Einstein equation
\begin{align} \label{eq: Einstein equation (00)+(0r)}
  g^{r0} (E_{00} - T_{00}) + g^{rr} (E_{r0} - T_{r0}) = 0
\end{align}
does not contain any scalar perturbations and it is called the first scalar constraint equation. To put \cref{eq: 1st order expanded metric} into the above one has
\begin{align}
  \frac1{r_H} \partial_0 r_H = - \frac2{9-p} \partial \beta. \label{eq: 1st order scalar constraint 1}
\end{align}
Since the above comes from the linear combination of $(00)$ and $(0r)$ components of Einstein equation. Both of them are independent of the spatial trace part of the metric. It is reasonable that \cref{eq: 1st order scalar constraint 1} contains no $q$.

A linear combination of $(0r)$ and $(rr)$ components of Einstein equation (which is also called the second scalar constraint equation)
\begin{align}
  g^{rr} (E_{rr} - T_{rr}) + g^{r0} (E_{r0} - T_{r0}) = 0, \label{eq: Einstein equation (rr)+(r0)}
\end{align}
the $(rr)$ component itself
\begin{align}
  E_{rr}-T_{rr} &= 0, \label{eq: Einstein equation (rr)}
\end{align}
and the EOM of $\phi$ \cref{eq: EOM dilaton} are chosen to solve the 3 scalar perturbations, whose first order differential equations are
\begin{align}
  & (r^{7-p} k_{(1)})' + 2 (7-p) r^{6-p} j_{(1)} - \left[ (p-q) r^{7-p} - \frac{2(p-q)}{9-p} r_H^{7-p} \right] h'_{(1)} - 2 r^\frac{7-p}{2} \partial \beta = 0, \label{eq: diff eq from 1st order scalar constraint2} \\
  & 2(p-q) r h''_{(1)} + (7-p)(p-q) h'_{(1)} - 2(9-p) j'_{(1)} = 0, \label{eq: diff eq from 1st order EOM(rr)} \\
  & 2(r^{7-p}k_{(1)})' + 2r^{7-p} f j'_{(1)} + 4(7-p) r^{6-p} j_{(1)} - (p-q) r^{7-p} f h'_{(1)} - 2r^\frac{7-p}2 \partial \beta = 0. \label{eq: diff eq from 1st order EOM dilaton}
\end{align}
Again, SO$(p-q)$ invariance allows us to set $\chi^{(1)} = F_\chi \partial \beta$ with $\chi = \{h,~j,~k\}$. The viscous term $\partial \beta$ can be removed:
\begin{align}
  & (r^{7-p} F_k)' + 2 (7-p) r^{6-p} F_j - \left[ (p-q) r^{7-p} - \frac{2(p-q)}{9-p} r_H^{7-p} \right] F_h' - 2 r^\frac{7-p}{2} = 0, \cr
  & 2(p-q) r F_h'' + (7-p)(p-q) F_h' - 2(9-p) F_j' = 0,  \cr
  & 2(r^{7-p}F_k)' + 2r^{7-p} f F_j' + 4(7-p) r^{6-p} F_j - (p-q) r^{7-p} f F_h' - 2r^\frac{7-p}2  = 0. \label{eq: diff eqns for F_h F_j and F_k}
\end{align}
The first order solutions for the 3 scalar perturbations are
\begin{align}
  F_h &= \frac1{p-q} F, \quad F_j = - \frac{2}{9-p} \frac{ r^\frac{9-p}{2} - r_H^\frac{9-p}{2} }{ r^{7-p} - r_H^{7-p} } + \frac{5-p}{2(9-p)} F, \cr
  F_k &= \frac4{ (9-p) r^\frac{5-p}{2} } - \frac{1}{9-p} \left( 5-p + \frac{2r_H^{7-p}}{r^{7-p}} \right) F. \label{eq: 1st order hjk solutions}
\end{align}
From the solution one can see that among the 3 first order scalar perturbations, only $h$ is affected by compactification. This is because $h$ is the spatial trace part of the metric perturbations. It is sensitive to the total number of the spatial dimensions and should be under affect of the compactification. From the view of boundary fluid, $h^{(1)}$ directly relates to the spatial trace part of $\partial_\mu u_\nu$, i.e. $\partial \cdot u$. Thus it is reasonable for $h^{(1)}$ to be dependent on the total number of the spatial dimensions. This is the reason that only $h$ is affected by the compactification among the 3 scalar perturbations. The factor $p-q$ in the denominator of the expression of $F_h$ is exactly the sign that the number of spatial directions of the compactified Dp-brane is now $p-q$ but not $p$ as in \cite{Wu1807}.

The differential equations of $F_j$ and $F_k$ do not depend on $q$. This can be seen by substituting $(p-q) F_h$ for $F$ in \cref{eq: diff eqns for F_h F_j and F_k}, then $(p-q)$ disappears from all the differential equations in \cref{eq: diff eqns for F_h F_j and F_k}. Then one can get the expressions for $F_k$ and $F_j$ in terms of $F$. In such way we do not need to solve $F_j$ and $F_k$ out explicitly, which is much more difficult than to find the relations directly as in \cref{eq: 1st order hjk solutions}.
The $q$-independence of $k$ and $j$ may also be figured out in the way that they are the $(00)$ and $(0r)$ components of the perturbative metric, respectively. Both of them have nothing to do with the number of the spatial dimensions. Compactification on spatial directions should not affect the scalar perturbations $k$ and $j$.

\subsection{Constitutive relation on the boundary}

The constitutive relation of the fluid living on the boundary can be got from taking large $r$ limit for the Brown-York tensor of the bulk theory:
\begin{align}\label{eq: Brown-York tensor}
  T_{\mu\nu} = \frac{1}{\kappa_{p-q+2}^2} \lim_{r\to\infty} \left( \frac r{L_p} \right)^\frac{(9-p)(p-q-1)}{2(p-q)} \left( K_{\mu\nu} - h_{\mu\nu}K - \frac{9-p}{2{L_p}} \left( \frac r{L_p} \right)^{-\frac{(p-3)^2 + q(5-p)}{2(p-q)}} h_{\mu\nu} \right),
\end{align}
The third term in the parenthesis of the above is from the counter term of the action. Note that the difference of the two powers of the factor $\frac{r}{L_p}$ is always 2 for all Dp-brane with or without compactification. Here
\begin{align}
  K_{\mu\nu} = - \frac12 \left( n^\rho \partial_\rho h_{\mu\nu} + \partial_\mu n^\rho h_{\rho\nu} + \partial_\nu n^\rho h_{\rho\mu} \right),
\end{align}
and $K = h^{\mu\nu} K_{\mu\nu}$. From the factor $\left( \frac r{L_p} \right)^\frac{(9-p)(p-q-1)}{2(p-q)}$ in \cref{eq: Brown-York tensor} we can see that compactification of the bulk gravity will change the asymptotic behavior of the Brown-York tensor in the framework of fluid/gravity correspondence.

Putting \cref{eq: 1st order expanded metric} into \cref{eq: Brown-York tensor} with all solved first order perturbations we then have the first order constitutive relation of the boundary fluid as
\begin{align}\label{eq: 1st order stress-energy tensor}
  T_{\mu\nu} =&\, \frac{1}{2 \kappa_{p-q+2}^2} \left[ {r_H^{7-p} \over L_p^{8-p}} \left( \frac{9-p}{2} u_\mu u_\nu + \frac{5-p}{2} P_{\mu\nu} \right) \right. \cr
  &\left. - \left( \frac{r_H}{L_p} \right) ^\frac{9-p}{2} \left( 2\sigma_{\mu\nu} + \frac{2(p-3)^2 + 2q(5-p)}{(p-q)(9-p)} P_{\mu\nu} \partial u \right) \right].
\end{align}
Then thermal and transport coefficients of first order can be got as
\begin{align}\label{eq: 1st order transport coefficients}
  \varepsilon &= \frac{1}{2 \kappa_{p-q+2}^2} \frac{9-p}{2} {r_H^{7-p} \over L_p^{8-p}}, \qquad \mathfrak p = \frac{1}{2 \kappa_{p-q+2}^2} \frac{5-p}{2} {r_H^{7-p} \over L_p^{8-p}}, \cr
   \eta &= \frac{1}{2 \kappa_{p-q+2}^2} \left( \frac{r_H}{L_p} \right)^\frac{9-p}{2}, \qquad \zeta = \frac{1}{2 \kappa_{p-q+2}^2} \frac{2(p-3)^2 + 2q(5-p)}{(p-q)(9-p)} \left( \frac{r_H}{L_p} \right)^\frac{9-p}{2}.
\end{align}
In the above, only the bulk viscosity depends on the number of compact directions $q$ since it relates with the spatial trace of the metric perturbation for which only $h^{(1)}=F_h \partial \beta$ contributes.

From these results, one can see that the ratio $\zeta/\eta$ that we get via fluid/gravity correspondence is exactly what is proposed in \cite{Mas0703} as quoted in \cref{eq: zeta over eta in Mas0703}. Thus the expression for the ratio $\zeta/\eta$ given in \cite{Mas0703} is indeed for compactified Dp-brane. The Hawking temperature $T$ for compactified Dp-brane is the same as Dp-brane. So the entropy density $s=(\varepsilon+p)/T$ as well as the ratio $\eta/s$ are also the same as that of Dp-brane case.

\section{Preliminaries on second order calculation} \label{sec: preliminaries on the 2nd order}

\subsection{The second order constraints and Navier-Stokes equations}

In the first order calculation we have two equations that without any perturbation, which are actually the Navier-Stokes equations of the boundary fluid at the first order. But from the gravity viewpoint they are the first order scalar \cref{eq: 1st order scalar constraint 1} and vector \cref{eq: 1st order vector constraint} constraint. They relate the following 5 first order partial derivative terms of the collective modes of different type:
\begin{align}
  \partial_0 r_H, \qquad \partial \beta, \qquad \partial_i r_H, \qquad \partial_0 \beta_i, \qquad \sigma_{ij} = \partial_{(i}\beta_{j)} - \frac1{p-q} \delta_{ij} \partial \beta.
\end{align}
Thus only three of the above are actually independent. We choose $\partial \beta$, $\partial_0 \beta_i$ and $\sigma_{ij}$ as the first order scalar, vector and tensor viscous term, respectively. That's why the first order perturbations can be written as $\chi^{(1)} = F_\chi \partial \beta$, $w^{(1)}_i = a(r) \partial_0 \beta_i$ and $\alpha_{ij}^{(1)} = F(r) \sigma_{ij}$.

The second order solving procedure follows the same rule. But now we have much more viscous terms since its complexity grows nonlinearly.
\begin{table}[h]
\centering
\begin{tabular}{|l|l|l|}
  \hline
  Scalars of $\mathrm{SO}(p-q)$                                 &                 Vectors of $\mathrm{SO}(p-q)$                      &\quad  Tensors of $\mathrm{SO}(p-q)$ \\ \hline\hline
 $\mathbf{s}_1=\frac1{r_H}\partial_0^2r_H$         & $\mathbf{v}_{1i} = \frac1{r_H} \partial_0\partial_i r_H$ & $\mathbf{t}_{1ij} = \frac1{r_H} \partial_i\partial_j r_H - \frac1{p-q} \delta_{ij} \mathbf{s}_3$ \\
  $\mathbf{s}_2 = \partial_0\partial_i\beta_i$           & $\mathbf{v}_{2i} = \partial_0^2\beta_i$       & $\mathbf{t}_{2ij} = \partial_0 \Omega_{ij}$ \\
  $\mathbf{s}_3 = \frac1{r_H}\partial_i^2r_H$         & $\mathbf{v}_{3i} = \partial_j^2\beta_i$        & $\mathbf{t}_{3ij} = \partial_0\sigma_{ij}$ \\
  $\mathfrak S_1 = \partial_0\beta_i\partial_0\beta_i$ & $\mathbf{v}_{4i}=\partial_j\Omega_{ij}$      & $\mathfrak T_{1ij} = \partial_0\beta_i\partial_0\beta_j - \frac1{p-q} \delta_{ij} \mathfrak S_1$ \\
  $\mathfrak S_3 = (\partial_i\beta_i)^2$                    & $\mathbf{v}_{5i} = \partial_j\sigma_{ij}$      & $\mathfrak T_{2ij} = \sigma_{[i}^{~~k} \Omega_{j]k}$ \\
  $\mathfrak S_4 = \Omega_{ij} \Omega_{ij}$           & $\mathfrak V_{1i} = \partial_0\beta_i\partial\beta$     & $\mathfrak T_{3ij} = \Omega_{ij} \partial \beta$ \\
  $\mathfrak S_5=\sigma_{ij}\sigma_{ij}$                 &  $\mathfrak V_{2i} = \partial_0\beta_j \Omega_{ij}$  & $\mathfrak T_{4ij}=\sigma_{ij}\partial\beta$ \\
                                                                                        & $\mathfrak V_{3i} = \partial_0\beta_j \sigma_{ij}$  & $\mathfrak T_{5ij} = \Omega_i^{~k}\Omega_{jk} - \frac1{p-q} \delta_{ij} \mathfrak S_4$ \\
                                                                                        &                                                             & $\mathfrak T_{6ij} = \sigma_i^{~k}\sigma_{jk} - \frac1{p-q} \delta_{ij} \mathfrak S_5$ \\
                                                                                        &                                                             & $\mathfrak T_{7ij} = \sigma_{(i}^{~~k} \Omega_{j)k}$ \\
\hline
\end{tabular}
\caption{\label{tab: 2nd order spatial viscous tensors} The list of $\mathrm{SO}(p-q)$ invariant second order viscous terms for $1\leq p-q \leq 3$. $\mathfrak T_{2,5,6}$ do not exist when $p-q=2$. At $p-q=1$, only $\mathbf s _{1,2,3},~\mathfrak S_{1,3}, ~\mathbf v_{1,2,3}$ and $\mathfrak V_1$ exist.}
\end{table}
These second order viscous terms satisfy 6 identities which can be extract from $\partial_\mu \partial^\rho T^{(0)}_{\rho\nu} = 0$. $T^{(0)}_{\mu\nu}$ here is the thermodynamical part of \cref{eq: 1st order stress-energy tensor} with $r_H$ and $u^\mu$ depend on $x^\mu$. After we expand $T^{(0)}_{\mu\nu}$ and take different components of $\partial_\mu \partial^\rho T^{(0)}_{\rho\nu} = 0$, it gives
\begin{align}
  \frac{9-p}{2} \frac{1}{r_H} \partial _0^2 r_H + \partial_0\partial \beta - \frac{2}{9-p} (\partial \beta)^2 - \frac{4}{5-p} \partial_0\beta_i \partial_0\beta_i = 0, \label{eq: 2nd order constraint (00) raw}  \\
  \frac{5-p}{2} \frac{1}{r_H} \partial_i^2 r_H + \partial_0 \partial \beta - \frac{2}{5-p} \partial_0\beta_i \partial_0\beta_i - \frac{5-p}{9-p} (\partial \beta)^2 + \partial_i \beta_j \partial _j\beta_i = 0, \label{eq: 2nd order constraint (ii) raw} \\
  \frac{5-p}{2} \frac{1}{r_H} \partial _0\partial _i r_H + \partial _0^2\beta_i - \frac{7-p}{9-p} \partial _0\beta_i \partial \beta + \partial_0 \beta_j \partial _j\beta_i = 0, \label{eq: 2nd order constraint (0i) raw} \\
  \frac{9-p}{2} \frac{1}{r_H} \partial _0\partial _i r_H + \partial_i \partial \beta - \frac{2}{5-p} \partial _0\beta_i \partial \beta - \frac{4}{5-p} \partial_0 \beta_j \partial _i\beta_j = 0, \label{eq: 2nd order constraint (i0) raw} \\
  \partial _0 \Omega_{ij} - \frac{5-p}{9-p} \Omega_{ij} \partial \beta - \partial_k\beta_{[i} \partial_{j]}\beta_k = 0, \label{eq: 2nd order constraint [ij] raw} \\
  \frac{5-p}{2} \frac{1}{r_H} \partial_i \partial _j r_H + \partial _0 \partial_{(i}\beta_{j)} - \frac{2}{5-p} \partial_0\beta_i \partial_0\beta_j - \frac{5-p}{9-p} \partial_{(i}\beta_{j)} \partial\beta + \partial _k\beta _{(i} \partial _{j)} \beta _k = 0. \label{eq: 2nd order constraint (ij) raw}
\end{align}
These are the same as the cases of Dp-brane without compact direction. But after reexpressed in terms of the spatial viscous tensors in \cref{tab: 2nd order spatial viscous tensors}, they actually depend on $q$:
\begin{align}
  & \mathbf s_1 + \frac2{9-p} \mathbf s_2 - \frac{8}{(9-p)(5-p)} \mathfrak S_1 - \frac{4}{(9-p)^2} \mathfrak S_3 = 0, \label{eq: 2nd order constraint (00)} \\
  & \mathbf s_2 + \frac{5-p}{2} \mathbf s_3 - \frac{2}{5-p} \mathfrak S_1 + \frac{(p-3)^2+q(5-p)}{(p-q)(9-p)} \mathfrak S_3 - \mathfrak
  S_4 + \mathfrak S_5 = 0, \label{eq: 2nd order constraint (ii)} \\
  & \mathbf v_1 + \frac{2}{5-p} \mathbf v_2 + \frac{2(p^2-8p+9 + q(7-p))}{(p-q)(9-p)(5-p)} \mathfrak V_1 - \frac{2}{5-p} \mathfrak V_2 + \frac{2}{5-p} \mathfrak V_3 = 0, \label{eq: 2nd order constraint (0i)} \\
  & \mathbf v_1 + \frac{2(p-q) (\mathbf v_4 + \mathbf v_5)}{(p-q-1)(9-p)} - \frac{4(p-q+2)}{(p-q)(9-p)(5-p)} \mathfrak V_1 - \frac{8 (\mathfrak V_2 + \mathfrak V_3)}{(9-p)(5-p)} = 0, \label{eq: 2nd order constraint (i0)} \\
  & \mathbf t_2 - 2 \mathfrak T_2 + \frac{p^2-7p+18 + q(5-p)}{(p-q)(9-p)} \mathfrak T_3 = 0, \label{eq: 2nd order constraint [ij]} \\
  & \mathbf t_1 + \frac{2}{5-p} \mathbf t_3 - \frac{4}{(5-p)^2} \mathfrak T_1 + \frac{2(p^2-7p+18) + 2q(5-p)}{(p-q)(9-p)(5-p)} \mathfrak T_4 - \frac{2}{5-p} \mathfrak T_5 + \frac{2}{5-p} \mathfrak T_6 = 0. \label{eq: 2nd order constraint (ij)}
\end{align}

The above 6 identities are used in deriving the differential equations for the second order perturbations as well as the second order Navier-Stokes equations. We will derive the latter in this section from a hydrodynamical viewpoint. They can be gained by expanding \cref{eq: 1st order stress-energy tensor} to the second order of partial derivatives and putting it into $\partial^\mu T^{(0+1)}_{\mu\nu} = 0$, where we add the superscript $(0+1)$ to emphasize the difference with $T^{(0)}_{\mu\nu}$. The second order Navier-Stokes equations are
\begin{align}
  \frac{1}{r_H^{(p-3)/2}} \partial_0 r_H^{(1)} =&\; \frac{4(p-3)^2 + 4q(5-p)}{(p-q)(9-p)^2(7-p)} \mathfrak S_3 + \frac{4}{(9-p)(7-p)} \mathfrak S_5, \label{eq: Navier-Stokes 0}\\
  \frac{1}{r_H^{(p-3)/2}} \partial_i r_H^{(1)} =&\; \frac{[4(p-3)^2 + 4q(5-p)]  \mathbf v_4 + 16(p-q)  \mathbf v_5}{(p-q-1)(9-p)(7-p)(5-p)} \cr
 & + \frac{2(p-1)(p^2 - 22p + 77) - 2q(p^2 - 22p + 85)}{(p-q)(9-p)(7-p)(5-p)^2} \mathfrak V_1 \cr
 & - \frac{2(19 - 3p)}{(9-p)(7-p)(5-p)} \mathfrak V_2 - \frac{2(p^2 - 14p + 77)}{(9-p)(7-p)(5-p)^2} \mathfrak V_3.
 \label{eq: Navier-Stokes i}
\end{align}
They are more complex than the Dp-brane case, when set $q=0$, they are back to the form of Dp-brane. The above second order Navier-Stokes equations can also be got through the constraints of Einstein equation \cref{eq: Einstein equation (00)+(0r),eq: Einstein eq (0i)+(ri)}, which will be specified later when we solve the second order perturbations.

\subsection{The second order expanded metric}

The calculation of the second order needs first to expand \cref{eq: x dependent global metric} to the second order of partial derivatives of $r_H$ and $\beta_i$, and the result is
\begin{align}\label{eq: 2nd order expanded metric}
  ds^2 =& -r^\frac{9-p}{p-q} \bigg[ f - (1-f) \delta\beta_i \delta\beta_i - \frac{(7-p)r_H^{6-p}}{r^{7-p}}\delta r_H - \frac{(7-p)r_H^{6-p}}{2r^{7-p}} \delta^2r_H - \frac{(7-p)r_H^{6-p}}{r^{7-p}}\delta r_H^{(1)} \cr
  & - \frac{(7-p)(6-p) r_H^{5-p}}{2r^{7-p}}(\delta r_H)^2 - (F_k+\delta F_k)\partial \beta - F_k(\delta\partial \beta + \delta\beta_i \partial_0\beta_i) - 2 a(r)\delta\beta_i \partial_0\beta_i \cr
  & - k^{(2)}(r) \bigg] dv^2 - 2r^\frac{9-p}{p-q} \bigg[ (1-f)(\delta\beta_i + \frac12 \delta^2\beta_i) + a(\partial_0\beta_i + \delta\partial_0\beta_i + \delta\beta_j\partial_j\beta_i) \cr
  & + \frac{(7-p)r_H^{6-p}}{r^{7-p}}\delta r_H \delta\beta_i + F_k \partial \beta \delta\beta_i + F \delta\beta_j \partial_{(i}\beta_{j)} + w_i^{(2)}(r) \bigg] dvdx^i \cr
  & + 2r^\frac{(p-3)(p-6) + q(7-p)}{2(p-q)} \bigg[ 1 + (F_j+\delta F_j) \partial \beta + F_j(\delta\partial \beta + \delta\beta_i\partial_0\beta_i) + \frac12 \delta\beta_i\delta\beta_i + j^{(2)}(r) \bigg] dvdr \cr
  & + r^\frac{9-p}{p-q} \bigg[ \delta_{ij} + (1-f)\delta\beta_i\delta\beta_j + 2a \delta\beta_{(i}\partial_{|0|}\beta_{j)} + (F + \delta F) \partial_{(i}\beta_{j)} + F \left( \delta\partial_{(i} \beta_{j)} + \delta\beta_{(i} \partial_{|0|}\beta_{j)} \right) \cr
  & + \alpha_{ij}^{(2)}(r) + h^{(2)}(r) \delta_{ij} \bigg] dx^idx^j - 2 r^\frac{(p-3)(p-6) + q(7-p)}{2(p-q)} \bigg( \delta\beta_i + \frac12 \delta^2\beta_i + F_j \partial\beta \delta\beta_i \bigg) dx^i dr.
\end{align}
Here we have $\delta^2 \psi = x^\mu x^\nu \partial_\mu \partial_\nu \psi$ with $\psi$ either $r_H$ or $\beta_i$, while
\begin{align}
  \delta \mathcal F(r_H(x),r) = - \frac{(5-p) \mathcal{F}(r) + 2r \mathcal{F}'(r)}{2r_H} \delta r_H
\end{align}
with $\mathcal{F}$ standing for any of the $F,~F_j$ and $F_k$. The above does not change compared with the Dp-brane case since $F,~F_j$ and $F_k$ do not change.

Each component of the above second order expanded metric has the structure of $g_{MN} = r^{(\cdots)}[\cdots]$, the expressions inside the brackets $[\cdots]$ do not change while terms inside the parenthesis $(\cdots)$ have changed compared with the Dp-brane case. The components of the second order expanded metric are now the second order polynomials of the boundary coordinates $x^\mu$ through terms like $\delta^2 \psi$ or $(\delta\psi)^2$ with $\psi$ being $r_H$ or $\beta_i$.

\section{Solving the second order perturbations} \label{sec: solve the 2nd order perturbations}

In the following subsections we will solve the second order perturbations for all the compactified Dp-branes with $p=2,~3,~4$ and $1\leq q \leq p-1$. The cases can be divided into three classes, in terms of $p-q$, they are $p-q=1,~2,~3$. Since $p-q=3$ is just the D4-brane with 1 direction compactified. We refer the reader to \cite{Wu1604} for details and will omit the solving procedure in this work despite the conventions used here is a little different from that of \cite{Wu1604}.

\subsection{The tensor part}

We will solve the second order tensor perturbations for all compactified Dp-brane in this subsection. Since the D(4-1)-brane case has been solved in \cite{Wu1604}, the cases contained here are only $p-q=2$ which include D(4-2) and D(3-1)-brane. The situation of $p-q=1$ should not be considered because the spatial viscous tensors can only be constructed in spacetime with more than 1 spatial dimensions.

The differential equation of the tensor perturbation is
\begin{align}\label{eq: diff eq of 2nd order alpha(ij) general form}
  \frac{d}{d r} \left( r^{8-p} f(r) \frac{d \alpha_{ij}^{(2)}}{d r} \right) = S_{ij}(r),
\end{align}
which is an exact second order linear inhomogeneous differential equation. The inhomogeneous term $S_{ij}(r)$ is called the source term in \cite{Bhattacharyya0712}. This equation is singular at $r=0$, $r=r_H$ and $r\to \infty$. Its solution can be got by integration twice as
\begin{align} \label{eq: integrate to get alpha}
  \alpha_{ij}^{(2)}(r) = \int_{r}^{\infty} \frac{-1}{x^{8-p} f}dx \int_{1}^{x} S_{ij}(y) dy,
\end{align}

We will solve the above equation by specifying $p,~q$ with explicit values. When putting the second order expanded metric \cref{eq: 2nd order expanded metric} into \cref{eq: Einstein eq(ij)} for D4-brane with 2 compact directions, one gets
\begin{align}\label{eq: D(4-2) diff eq for 2nd order alpha(ij)}
  \partial_r (r^4 f \partial_r \alpha_{ij}^{(2)}) =& \left( 6r - \frac52 r^\frac32 F - 2r^\frac52 F' \right) \left( \mathbf t_3 + \mathfrak T_1 \right) + \bigg[ 6r - \frac12 r^\frac32 F - \frac{21}5 r^{\frac52} F'  \cr
  & - \frac45 r^\frac72 F'' + \frac12 r^4 f F'^2 + (4r^3-1) FF' + r^4 f FF'' + 5 r^\frac32 F_j - 15 r^2 FF_j  \cr
  & + 2(4r^3-1) F' F_j - (5r^3-2) F F_j' + r^4 f F' F_j' + 2 r^4 f F'' F_j  \cr
  & - \frac{15}{2} r^2 FF_k + 4 r^3 F'F_k - \frac{13}{2} r^3 FF_k' + r^4 F' F_k' + r^4 F'' F_k - r^4 F F_k'' \bigg] \mathfrak T_4 \cr
  & - \left( 4r + 5r^\frac32 F + 4 r^\frac52 F' \right) \mathfrak T_7,
\end{align}
The source term on the right hand side of the above equation has the same structure as D2-brane. The differences lie only in the coefficient functions of each spatial viscous tensors. Using \cref{eq: integrate to get alpha}, one has
\begin{align}
  \alpha_{ij}^{(2)} =& \left( \frac43 - \frac{\pi}{9\sqrt3} - \frac{\ln3}{3} \right) \frac{1}{r^3} \left( t_3 + \mathfrak T_1 \right) + \left[ \frac{4}{r} + \left( \frac{16}{15} - \frac{\pi}{45\sqrt3} - \frac{\ln3}{15} \right) \frac{1}{r^3} \right] \mathfrak T_4 \cr
  & + \left[ \frac8r - \left( \frac{2\pi}{9\sqrt3} + \frac{2\ln3}{3} \right) \frac{1}{r^3} \right] \mathfrak T_7
\end{align}
This solution is in a form of an asymptotic series of $r$. It is got by making a series expansion in between the two sequential integrations for $y$ and $x$, as has been specified in \cite{Wu1604}. Expanding the source term at the beginning will lead to wrong results for $\alpha_{ij}^{(2)}$ and $h^{(2)}$. But this will not affect the integration for $w_i^{(2)}$, $j^{(2)}$ and $k^{(2)}$. We think the reason is that both the integrations for $\alpha_{ij}^{(2)}$ and $h^{(2)}$ depend on the emblackening factor $f(r)$ while $w_i^{(2)}$ do not. As for $j^{(2)}$ and $k^{(2)}$, they are calculated by integration only once. This exchangeability between the integral of $y$ and the asymptotic expansion will give us great help for the calculations.

Though near extremal D3-brane can be reduced trivially to AdS$_5$ black hole whose dual fluid is conformal. After compactification, it will correspond to nonconformal fluid. The tensor perturbation for D(3-1)-brane satisfies the differential equation
\begin{align}\label{eq: D(3-1) diff eq of 2nd order alpha}
  \partial_r (r^5 f \partial_r \alpha_{ij}^{(2)}) =& \left( 2r - 3 r^2 F - 2 r^3 F' \right) \left( \mathbf t_3 + \mathfrak T_1 \right) + \bigg[ 2r - r^2 F - \frac{13}3 r^3 F' - \frac23 r^4 F'' \cr
  & + (5r^4-1) FF' + r^5 f FF'' + \frac12 r^5 f F'^2 + 6 r^2 F_j - 24 r^3 FF_j  \cr
  & + 2(5r^4-1) F' F_j - 2(3r^4-1) F F_j' + r^5 f F' F_j' + 2 r^5 f F'' F_j  \cr
  & - 12 r^3 FF_k + 5 r^4 F' F_k - 8 r^4 F F_k' + r^5 F' F_k' + r^5 F'' F_k - r^5 F F_k'' \bigg] \mathfrak T_4 \cr
  & - \left( 4r + 6 r^2 F + 4 r^3 F' \right) \mathfrak T_7,
\end{align}
using \cref{eq: integrate to get alpha} one gets the solution as
\begin{align}
  \alpha_{ij}^{(2)} &= \left( \frac12 - \frac{\ln2}{4} \right) \frac{1}{r^4} (\mathbf t_3 + \mathfrak T_1) + \left[ \frac{1}{r^2} + \left( \frac13 - \frac{\ln2}{12} \right) \frac{1}{r^4} \right] \mathfrak T_4 + \left( \frac{2}{r^2} - \frac{\ln2}{2r^4} \right) \mathfrak T_7.
\end{align}
Compared with D3-brane case where the coefficient function of $\mathbf t_3 + \mathfrak T_1$ is proportional to that of $\mathfrak T_4$, these two coefficient functions here are not proportional to each other, suggesting the background is not conformal now.

From the calculation of this subsection, the $1/r$ dependent behavior of the tensor perturbations of D(p-q)-brane does not change compared with that of Dp-brane. Perturbations of other types share the same feature as the tensor perturbation, as can be seen from the solutions of vector and scalar perturbations. This shows that the compactification will not affect the asymptotic behavior of the metric perturbations. Actually, the asymptotic behavior of the metric perturbations are mainly determined by the power of $r$ in the emblackening factor $f(r)$ which is independent of the compact directions $q$.

\subsection{The vector part}

There are a constraint plus a dynamical equation for the vector part. Both the cases of $p-q=$ 2 and 1 need to be analyzed here.

\subsubsection{The constraint equation}

The vector constraint equation can be got by feeding \cref{eq: 2nd order expanded metric} into \cref{eq: Einstein eq (0i)+(ri)}. The result is just the spatial component of the Navier-Stokes equation of second order, as one can check by substituting specific values for $p$ and $q$ into \cref{eq: Navier-Stokes i}.

Let's start with $p-q=2$ first. For D(4-2)-brane, the vector constraint equation is
\begin{align}\label{eq: D(4-2) 2nd order vec constraint eq}
  \partial _i r_H^{(1)} &= \left[ -\frac43 r^\frac52 - \frac23 r^4 f F' + \frac43 (5r^3-2) F_j + \frac{10}3 r^3 F_k + \frac43 r^4 F_k' \right] \mathbf v_4 \cr
  & + \left[ \frac43 (5r^3-2) F_j + \frac{10}3 r^3 F_k + \frac43 r^4 F_k' \right] \mathbf v_5 \cr
  & + \left[ \frac{13}{3} r^\frac52 + \frac16 r (13 r^3-1) F' - (15r^3+2) F_j - \frac{15}{2} r^3 F_k - \frac{7}3 r^4 F_k' \right] \mathfrak V_1 \cr
  & + \left[ - 2 r^\frac52 - r^4 f F' + \frac23 (5r^3-2) F_j + \frac53 r^3 F_k + \frac23 r^4 F_k' \right] \mathfrak V_2 \cr
  & + \left[ - 6 r^\frac52 - 3r^4 f F' + \frac23 (5r^3-2) F_j + \frac53 r^3 F_k + \frac23 r^4 F_k' \right] \mathfrak V_3.
\end{align}
After expansion with respect to $1/r$, we have
\begin{align}
  \partial_i r_H^{(1)} = \frac{4}{5} \mathbf v_4 + \frac{32}{15} \mathbf v_5 - \frac{11}{15} \mathfrak V_1 - \frac{14}{15} \mathfrak V_2 - \frac{74}{15} \mathfrak V_3
\end{align}
By the same token, one has the vector constraint for D(3-1)-brane as
\begin{align}\label{eq: D(3-1) 2nd order vec constraint eq}
  \partial _i r_H^{(1)} &= \left[ -\frac12 r^3 - \frac14 r^5 f F' + (3r^4-1) F_j + \frac32 r^4 F_k + \frac12 r^5 F_k' \right] \mathbf v_4 \cr
  & + \left[ (3r^4-1) F_j + \frac32 r^4 F_k + \frac12 r^5 F_k' \right] \mathbf v_5 \cr
  & + \left[ \frac98 r^3 + \frac1{16} r (9 r^4-1) F' - \frac34 (5r^4+1) F_j - \frac{15}{8} r^4 F_k - \frac38 r^5 F_k' \right] \mathfrak V_1 \cr
  & + \left[ - \frac34 r^3 - \frac38 r^5 f F' + \frac12 (3r^4-1) F_j + \frac34 r^4 F_k + \frac14 r^5 F_k' \right] \mathfrak V_2 \cr
  & + \left[ - \frac54 r^3 - \frac58 r^5 f F' + \frac12 (3r^4-1) F_j + \frac34 r^4 F_k + \frac14 r^5 F_k' \right] \mathfrak V_3.
\end{align}
After expansion, the above becomes
\begin{align}
  \partial_i r_H^{(1)} = \frac16 \mathbf v_4 + \frac23 \mathbf v_5 + \frac18 \mathfrak V_1 - \frac{5}{12} \mathfrak V_2 - \frac{11}{12} \mathfrak V_3.
\end{align}

The above two cases have two spatial directions for the dual fluid. Next we consider situations for $p-q=1$, the vector viscous term in this case are only $\mathbf v_{1,2,3}$ and $\mathfrak V_1$. Thus the equations will turn out to be much simple now. In the order of D(4-3), D(3-2) and D(2-1)-brane, the vector constraint equations can be read as
\begin{align}
  \partial _i r_H^{(1)} =& \left[ \frac23 (5r^3-2) F_j + \frac53 r^3 F_k + \frac23 r^4 F_k' \right] \mathbf v_3 + \left[ \frac43 r^\frac52 + \frac23 r (r^3+2) F' \right. \cr
  &\left. - \frac83 (5r^3+1) F_j - \frac{20}3 r^3 F_k - 2 r^4 F_k' \right] \mathfrak V_1, \label{eq: D(4-3) 2nd order vec constraint eq} \\
  \partial _i r_H^{(1)} =& \left[ \frac12 (3r^4-1) F_j + \frac34 r^4 F_k + \frac14 r^5 F_k' \right] \mathbf v_3 + \left[ \frac12 r^3 + \frac14 r (r^4+1) F' \right. \cr
  &\left. - (3r^4+1) F_j - \frac32 r^4 F_k - \frac14 r^5 F_k' \right] \mathfrak V_1, \label{eq: D(3-2) 2nd order vec constraint eq} \\
  \partial _i r_H^{(1)} =& \left[ \frac2{15} (7r^5-2) F_j + \frac7{15} r^5 F_k + \frac2{15} r^6 F_k' \right] \mathbf v_3 + \left[ \frac4{15} r^\frac72 + \frac2{45} r (3r^5+2) F' \right. \cr
  &\left. - \frac8{45} (7r^5+3) F_j - \frac{28}{45} r^5 F_k - \frac{2}{45} r^6 F_k' \right] \mathfrak V_1, \label{eq: D(2-1) 2nd order vec constraint eq}
\end{align}
which, after expansion with respect to $1/r$, lead to
\begin{align}
  \partial_i r_H^{(1)} &= \frac{16}{15} \mathbf v_3 - \frac{16}{5} \mathfrak V_1, \qquad  \text{D(4-3)-brane} \\
  \partial_i r_H^{(1)} &= \frac13 \mathbf v_3 - \frac13 \mathfrak V_1, \qquad \text{D(3-2)-brane} \\
  \partial_i r_H^{(1)} &= \frac{16}{105} \mathbf v_3 - \frac{16}{315} \mathfrak V_1. \qquad \text{D(2-1)-brane}
\end{align}

\subsubsection{The dynamical equation}

The differential equation of $w_i^{(2)}$ can be got by plugging the second order expanded metric \cref{eq: 2nd order expanded metric} into \cref{eq: Einstein eq(ri)}:
\begin{align}
  \frac{d}{d r} \left( r^{8-p} \frac{d w_i^{(2)}}{d r} \right) = S_i (r),
\end{align}
where $S_i$ in the right hand side is the source term. This equation is similar like \cref{eq: diff eq of 2nd order alpha(ij) general form}, the only difference is that the above does not have the emblackening factor in the left hand side, thus no singularity at $r=r_H$. The above equation can be solved by direct integration as
\begin{align}
  w_{i}^{(2)} &= \int_{r}^{\infty} \frac{1}{x^{8-p}} dx \int_{x}^{\infty} S_i(y) dy. \label{eq: integrate to get w}
\end{align}

Let's solve the $p-q=2$ cases first. For D(4-2)-brane, the differential equation of $w_i^{(2)}$ is
\begin{align}\label{eq: D(4-2) diff eq of 2nd order w}
  \partial_r(r^4\partial_r w_i^{(2)}) =&\; \left( - 2r - r^\frac52 F' + 5 r^\frac32 F_j - 2 r^\frac52 F_j' \right) \mathbf v_4 + \left(  5 r^\frac32 F_j - 2 r^\frac52 F_j' \right) \mathbf v_5 \cr
  & + \left( \frac12 r - \frac{19}4 r^\frac52 F' - 2 r^\frac72 F'' + \frac{25}{4} r^\frac32 F_j + \frac{5}{2} r^\frac52 F_j' - 2 r^\frac72 F_j'' \right) \mathfrak V_1 \cr
  & + \left( 5r - \frac32 r^\frac52 F' + \frac52 r^\frac32 F_j - r^\frac52 F_j' \right) \mathfrak V_2 \cr
  & + \left( -r - \frac92 r^\frac52 F' + \frac52 r^\frac32 F_j - r^\frac52 F_j' \right) \mathfrak V_3,
\end{align}
of which the solution is very simple
\begin{align}
  w_i^{(2)} = - \frac2{r} \mathfrak V_1 - \frac4r \mathfrak V_2 - \frac4r \mathfrak V_3.
\end{align}
The vector dynamical equation for the second order perturbation of D(3-1)-brane is
\begin{align}\label{eq: D(3-1) diff eq of 2nd order w}
  \partial_r (r^5 \partial_r w_i^{(2)}) =&\; \left( - 2r - r^3 F' + 6 r^2 F_j - 2 r^3 F_j' \right) \mathbf v_4 + \left(  6 r^2 F_j - 2 r^3 F_j' \right) \mathbf v_5 \cr
  & + \left( \frac12 r - \frac{11}4 r^3 F' - r^4 F'' + \frac{15}{2} r^2 F_j + \frac32 r^3 F_j' - r^4 F_j'' \right) \mathfrak V_1 \cr
  & + \left( r - \frac32 r^3 F' + 3 r^2 F_j - r^3 F_j' \right) \mathfrak V_2 \cr
  & + \left( -r - \frac52 r^3 F' + 3 r^2 F_j - r^3 F_j' \right) \mathfrak V_3,
\end{align}
with the solution is
\begin{align}
  w_i^{(2)} = - \frac1{2r^2} \mathfrak V_1 - \frac1{r^2} \mathfrak V_2 - \frac1{r^2} \mathfrak V_3.
\end{align}

For the cases of $p-q=1$, the differential equation for $w_i^{(2)}$ becomes simple again. In the order of D(4-3), D(3-2) and D(2-1)-brane, they are
\begin{align}
  \partial_r (r^4\partial_r w_i^{(2)}) &= \left(  \frac52 r^\frac32 F_j - r^\frac52 F_j' \right) \mathbf v_3 + \left( - 7 r^\frac52 F' - 2 r^\frac72 F'' + \frac{15}2 r^\frac32 F_j + 2 r^\frac52 F_j' - 2 r^\frac72 F_j'' \right) \mathfrak V_1, \label{eq: D(4-3) 2nd order vec dynamical eq} \\
   \partial_r (r^5 \partial_r w_i^{(2)}) &= \left( 3 r^2 F_j - r^3 F_j' \right) \mathbf v_3 + \left( - 4 r^3 F' - r^4 F'' + 9 r^2 F_j + r^3 F_j' - r^4 F_j'' \right) \mathfrak V_1, \label{eq: D(3-2) 2nd order vec dynamical eq} \\
   \partial_r (r^6 \partial_r w_i^{(2)}) &= \left(  \frac72 r^\frac52 F_j - r^\frac72 F_j' \right) \mathbf v_3 + \left( - 3 r^\frac72 F' - \frac23 r^\frac92 F'' + \frac{21}2 r^\frac52 F_j + \frac23 r^\frac72 F_j' - \frac23 r^\frac92 F_j'' \right) \mathfrak V_1. \label{eq: D(2-1) 2nd order vec dynamical eq}
\end{align}
The above 3 equations can also be identified by the power of $r$ in the left hand side: this power is $8-p$ for D(p-q)-brane. The solutions of the above are
\begin{align}
  w_i^{(2)} = - \frac4{r} \mathfrak V_1, & \qquad \text{D(4-3)-brane}\\
  w_i^{(2)} = - \frac1{r^2} \mathfrak V_1, & \qquad \text{D(3-2)-brane} \\
  w_i^{(2)} = - \frac4{9r^3} \mathfrak V_1. & \qquad \text{D(2-1)-brane}
\end{align}

\subsection{The scalar part}

The scalar perturbations play an important role in the nonconformal background. They are also more difficult to solve than the tensor and vector part because the scalar perturbations are mixed together in the differential equations derived from scalar components of Einstein equation and the EOM of dilaton (or the scalar $A$).

The differential equation of $h^{(2)}$ for D(p-q)-brane is
\begin{align}
  \frac{d}{d r} \left( r^{8-p} f(r) \frac{d h^{(2)}}{d r} \right) = S_h(r),
\end{align}
whose solution can be written as
\begin{align} \label{eq: integrate to get h}
  h^{(2)}(r) = \int_{r}^{\infty} \frac{-1}{x^{8-p} f}dx \int_{1}^{x} S_h(y) dy.
\end{align}
This is very like $\alpha_{ij}^{(2)}$, which can be seen from \cref{eq: diff eq of 2nd order alpha(ij) general form,eq: integrate to get alpha}.

The differential equations for $j$ and $k$ are actually first order ones thus they can be solved by integrating once as
\begin{align}
   j^{(2)}(r) &= - \int_r^\infty S_j(x) dx, \\
  k^{(2)}(r) &= - \frac{1}{r^{7-p}} \int_r^\infty S_k(x) dx.
\end{align}
$\frac{1}{r^{7-p}}$ is present in from of the integration. This is because $k$ always appear as $(r^{7-p}k)$ in the differential equations.

\subsubsection{The first scalar constraint}

We still first deal with the cases of $p-q=2$ in the scalar sector. The first scalar constraint does not contain any unknown perturbation and it can be calculated by plugging the second order expanded metric into \cref{eq: Einstein equation (00)+(0r)}, which gives for the D(4-2)-brane case as
\begin{align}\label{eq:  D(4-2) 2nd order scalar constraint 1}
  \partial _0 r_H^{(1)} &= \left( \frac{4}{5r^\frac12} + \frac{4}{15} r^\frac52 - \frac15 F + \frac{2}{15} r^4 f F' - \frac23 r^3 f F_j - \frac13 r^3 F_k \right) \mathbf s_2 + \frac{2}{5r^\frac12} \mathbf s_3 \cr
  & + \left( - \frac{8}{5r^\frac12} + \frac{4}{15}r^\frac52 - \frac15 F + \frac{2}{15} r^4 f F' - \frac23 r^3 f F_j - \frac13 r^3 F_k  \right) \mathfrak S_1 \cr
  & + \left( \frac{6}{25 r^\frac12} + \frac{2}{15} r^\frac52 - \frac{1}{25} F + \frac{11}{75} r^4 f F' + \frac{4}{75} r^5 f F'' - \frac{2}{15} r^3 f F_j - \frac{4}{15} r^4 f F_j' \right. \cr
  & \left. - \frac{1}{15} r^3 F_k \right) \mathfrak S_3 - \frac{4}{5 r^\frac12} \mathfrak S_4 + \left( \frac{4}{5 r^\frac12} + \frac{4}{15} r^\frac52 + \frac{2}{15} r^4 f F' \right) \mathfrak S_5
\end{align}
After expanding in powers of $1/r$, one has
\begin{align}
  \partial _0 r_H^{(1)} &= \frac{2}{25} \mathfrak S_3 + \frac{4}{15} \mathfrak S_5 + \frac{4}{5r^{1/2}} \left( \mathbf s_2 + \frac12 \mathbf s_3 - 2\mathfrak S_1 + \frac{3}{10} \mathfrak S_3 - \mathfrak S_4 + \mathfrak S_5 \right).
\end{align}
The expression inside the parenthesis is just the identity \cref{eq: 2nd order constraint (ii)} that satisfied by the second order spatial viscous terms thus gives a zero. So the above finally becomes
\begin{align}
  \partial _0 r_H^{(1)} &= \frac{2}{25} \mathfrak S_3 + \frac{4}{15} \mathfrak S_5
\end{align}
which is actually \cref{eq: Navier-Stokes 0} in the case of $p=4$, $q=2$.

\cref{eq: Einstein equation (00)+(0r)} gives us for D(3-1)-brane the first scalar constraint as
\begin{align}\label{eq: D(3-1) 2nd order scalar constraint 1}
  \partial _0 r_H^{(1)} &= \left( \frac1{3r} + \frac16 r^3 - \frac16 F + \frac1{12} r^5 f F' - \frac12 r^4 f F_j - \frac14 r^4 F_k \right) \mathbf s_2 + \frac1{3r} \mathbf s_3 \cr
  & + \left( - \frac1{3r} + \frac16 r^3 - \frac16 F + \frac1{12} r^5 f F' - \frac12 r^4 f F_j - \frac14 r^4 F_k  \right) \mathfrak S_1 \cr
  & + \left( \frac1{18 r} + \frac1{12} r^3 - \frac{1}{18} F + \frac{7}{72} r^5 f F' + \frac1{36} r^6 f F'' - \frac16 r^4 f F_j - \frac16 r^5 f F_j' \right. \cr
  & \left. - \frac{1}{12} r^4 F_k \right) \mathfrak S_3 - \frac1{3 r} \mathfrak S_4 + \left( \frac1{3 r} + \frac16 r^3 + \frac1{12} r^5 f F' \right) \mathfrak S_5,
\end{align}
which gives
\begin{align}
  \partial _0 r_H^{(1)} &= \frac{1}{36} \mathfrak S_3 + \frac{1}{6} \mathfrak S_5 + \frac{1}{3r} \left( \mathbf s_2 + \mathbf s_3 - \mathfrak S_1 + \frac16 \mathfrak S_3 - \mathfrak S_4 + \mathfrak S_5 \right).
\end{align}
Again, terms in the parenthesis sum to zero by \cref{eq: 2nd order constraint (ii)} and one finally has the scalar component of Navier-Stokes equation as
\begin{align}
  \partial _0 r_H^{(1)} &= \frac{1}{36} \mathfrak S_3 + \frac{1}{6} \mathfrak S_5.
\end{align}

Then we solve the situations of $p-q=1$. The first scalar constraint for D(4-3), D(3-2) and D(2-1)-brane can be listed separately as
\begin{align}
  \partial _0 r_H^{(1)} &= \left( \frac{4}{5 r^\frac12} + \frac{4}{15} r^\frac52 - \frac15 F + \frac{2}{15} r^4 f F' - \frac23 r^3 f F_j - \frac13 r^3 F_k \right) \mathbf s_2 + \frac{2}{5r^\frac12} \mathbf s_3 \cr
  & + \left( - \frac{8}{5r^\frac12} + \frac{4}{15} r^\frac52 - \frac15 F + \frac{2}{15} r^4 f F' - \frac23 r^3 f F_j - \frac13 r^3 F_k  \right) \mathfrak S_1 \cr
  & + \left( \frac{16}{25 r^\frac12} + \frac{4}{15} r^\frac52 - \frac{1}{25} F + \frac{16}{75} r^4 f F' + \frac{4}{75} r^5 f F'' - \frac{2}{15} r^3 f F_j - \frac{4}{15} r^4 f F_j' \right. \cr
  & \left. - \frac{1}{15} r^3 F_k \right) \mathfrak S_3, \label{eq:  D(4-3) 2nd order scalar constraint 1} \\
  \partial _0 r_H^{(1)} &= \left( \frac1{3 r} + \frac16 r^3 - \frac16 F + \frac1{12} r^5 f F' - \frac12 r^4 f F_j - \frac14 r^4 F_k \right) \mathbf s_2 + \frac1{3r} \mathbf s_3 \cr
  & + \left( - \frac1{3r} + \frac16 r^3 - \frac16 F + \frac1{12} r^5 f F' - \frac12 r^4 f F_j - \frac14 r^4 F_k  \right) \mathfrak S_1 \cr
  & + \left( \frac2{9 r} + \frac16 r^3 - \frac{1}{18} F + \frac{5}{36} r^5 f F' + \frac1{36} r^6 f F'' - \frac16 r^4 f F_j - \frac16 r^5 f F_j' \right. \cr
  & \left. - \frac{1}{12} r^4 F_k \right) \mathfrak S_3, \label{eq:  D(3-2) 2nd order scalar constraint 1} \\
  \partial _0 r_H^{(1)} &= \left( \frac{4}{21 r^\frac32} + \frac{4}{35} r^\frac72 - \frac17 F + \frac{2}{35} r^6 f F' - \frac25 r^5 f F_j - \frac15 r^5 F_k \right) \mathbf s_2 + \frac{2}{7 r^\frac32} \mathbf s_3 \cr
  & + \left( - \frac{8}{63 r^\frac32} + \frac{4}{35} r^\frac72 - \frac17 F + \frac{2}{35} r^6 f F' - \frac25 r^5 f F_j - \frac15 r^5 F_k  \right) \mathfrak S_1 \cr
  & + \left( \frac{16}{147 r^\frac32} + \frac{4}{35} r^\frac72 - \frac{3}{49} F + \frac{24}{245} r^6 f F' + \frac{4}{245} r^7 f F'' - \frac{6}{35} r^5 f F_j - \frac{4}{35} r^6 f F_j' \right. \cr
  & \left. - \frac{3}{35} r^5 F_k \right) \mathfrak S_3. \label{eq:  D(2-1) 2nd order scalar constraint 1}
\end{align}
After expansion in terms of powers of inverse $r$, the above become
\begin{align}
  \partial _0 r_H^{(1)} &= \frac{16}{75} \mathfrak S_3 + \frac{4}{5r^{1/2}} \left( \mathbf s_2 + \frac12 \mathbf s_3 - 2\mathfrak S_1 + \frac{4}{5} \mathfrak S_3 \right), \qquad \text{D(4-3)-brane} \\
  \partial _0 r_H^{(1)} &= \frac19 \mathfrak S_3 + \frac1{3r} \left( \mathbf s_2 + \mathbf s_3 - \mathfrak S_1 + \frac23 \mathfrak S_3 \right), \qquad \text{D(3-2)-brane} \\
  \partial _0 r_H^{(1)} &= \frac{16}{245} \mathfrak S_3 + \frac{4}{21r^{3/2}} \left( \mathbf s_2 + \frac32 \mathbf s_3 - \frac23 \mathfrak S_1 + \frac47 \mathfrak S_3 \right). \qquad \text{D(2-1)-brane}
\end{align}
Once more, terms in the parenthesis of the above are all zero in the name of \cref{eq: 2nd order constraint (ii)}. Thus one has finally
\begin{align}
  \partial _0 r_H^{(1)} &= \frac{16}{75} \mathfrak S_3, \qquad \text{D(4-3)-brane} \\
  \partial _0 r_H^{(1)} &= \frac19 \mathfrak S_3, \qquad \text{D(3-2)-brane} \\
  \partial _0 r_H^{(1)} &= \frac{16}{245} \mathfrak S_3. \qquad \text{D(2-1)-brane}
\end{align}

\subsubsection{The scalar dynamical equations}

To solve the 3 scalar perturbations, we need the second scalar constraint \cref{eq: Einstein equation (rr)+(r0)}, the $(rr)$ component of Einstein equation \cref{eq: Einstein equation (rr)} and the EOM of dilaton \cref{eq: EOM dilaton} or the scalar $A$ \cref{eq: EOM A}. Note the compactified D3-brane does not have dilaton field in the background, so we need EOM of $A$ to solve the scalar perturbations. For compactified D4 and D2-brane, one can just use EOM of dilaton.

We still tackle the cases of $p-q=2$ first. In the case of D(4-2)-brane, after feeding the equations mentioned above with \cref{eq: 2nd order expanded metric}, we get
\begin{align}\label{eq: D(4-2) 2nd order scalar constraint 2}
  & 5(r^3 k_{(2)})' + 30 r^2 j_{(2)} - 2(5r^3-2) h_{(2)}' = \left( -16r + 5 r^\frac32 F + 2 r^\frac52 F' \right) \mathbf s_2 \cr
  & + \left( 24r + 5r^\frac32 F + 2r^\frac52 F' \right) \mathfrak S_1 + \bigg[ - 7r + r^\frac32 F + \frac{26}{5} r^\frac52 F' + \frac45 r^\frac72 F'' - \frac12 (5r^3-2) FF' \cr
  & + \frac14 r^4 f F'^2 - 10 r^\frac32 F_j + 45 r^2 F_j^2 - 2(5r^3-2) F'F_j - 5r^3 F'F_k - r^4 F' F_k' + 30 r^2 F_j F_k \cr
  & + 10 r^3 F_j F_k' \bigg] \mathfrak S_3 + \left( \frac{2}{r^2} + 20r \right) \mathfrak S_4 - \left[ 18r + (5r^3-2)FF' + \frac12 r^4 f F'^2 \right] \mathfrak S_5,
\end{align}
\begin{align}
  2r h_{(2)}'' + 3 h_{(2)}' - 5 j_{(2)}' =& \left( \frac34 FF' + \frac12 r FF'' + \frac14 r F'^2 + r F' F_j' - 5 F_j F_j' \right) \mathfrak S_3 \cr
  & + \frac{2}{r^2} \mathfrak S_4 + \left( \frac32 FF' + r FF'' + \frac12 r F'^2 \right) \mathfrak S_5,
\end{align}
and
\begin{align}
  & (r^3 k_{(2)})' + r^3 f j_{(2)}' + 6 r^2 j_{(2)} - r^3 f h_{(2)}' = \left( -2r + \frac12 r^\frac32 F \right) \mathbf s_2 + \left( 6r + \frac12 r^\frac32 F \right) \mathfrak S_1   \cr
  & + \bigg[ - r + \frac{1}{10} r^\frac32 F + \frac15 r^\frac52 F' - \frac14 r^3 f FF' - r^\frac32 F_j + 9 r^2 F_j^2 - r^3 f F' F_j + 3 r^3 f F_j F_j'  \cr
  & - \frac12 r^3 F' F_k + 6 r^2 F_j F_k + r^3 F_j' F_k + 2 r^3 F_j F_k'  \bigg] \mathfrak S_3 + 2r \mathfrak S_4 - \left( 2r + \frac12 r^3 f FF' \right) \mathfrak S_5,
\end{align}
respectively. Then the perturbations can be solved as
\begin{align}
  h ^{(2)} =& \left( \frac23 - \frac{\pi}{18\sqrt3} - \frac{\ln3}{6} \right) \frac{1}{r^3} ( \mathbf s_2 + \mathfrak S_1 ) + \left[ \frac{1}{r} + \left( \frac15 - \frac{\pi}{90\sqrt3} - \frac{\ln3}{30} \right)\frac{1}{r^3} \right] \mathfrak S_3 \cr
  & + \frac2{r} \mathfrak S_4 + \left( \frac2{r} + \frac{2}{3r^3} \right) \mathfrak S_5, \\
  j^{(2)} =& - \left( \frac23 - \frac{\pi}{18\sqrt3} - \frac{\ln3}{6} \right) \frac{1}{r^3} ( \mathbf s_2 + \mathfrak S_1 ) - \left( \frac{1}{5} - \frac{\pi}{90\sqrt3} - \frac{\ln3}{30} \right) \frac{1}{r^3} \mathfrak S_3 - \frac{2}{3r^3} \mathfrak S_5, \\
  k^{(2)} =& \left[ - \left( \frac{4}{15} - \frac{\pi}{45\sqrt3} - \frac{\ln3}{15} \right) \frac{1}{r^3} + \frac{1}{10 r^4} - \left( \frac23 - \frac{\pi}{18\sqrt3} - \frac{\ln3}{6} \right)\frac{1}{r^6} \right] \mathbf s_2 \cr
  & + \left[ \frac{4}{r} - \left( \frac{4}{15} - \frac{\pi}{45\sqrt3} - \frac{\ln3}{15} \right) \frac{1}{r^3} + \frac{1}{10r^4} - \left( \frac23 - \frac{\pi}{18\sqrt3} - \frac{\ln3}{6} \right) \frac{1}{r^6} \right] \mathfrak S_1 \cr
  & + \left[ - \left( \frac{2}{25} - \frac{\pi}{225\sqrt3} - \frac{\ln3}{75} \right) \frac{1}{r^3} + \frac{12}{35 r^{7/2}} + \frac{47}{25 r^4} - \left( \frac{7}{45} - \frac{\pi}{90\sqrt3} - \frac{\ln3}{30} \right) \frac{1}{r^6} \right] \mathfrak S_3 \cr
  & - \frac2{r^4} \mathfrak S_4 - \left( \frac{4}{15 r^3} - \frac{8}{7 r^{7/2}} - \frac1{r^4} + \frac{2}{3r^6} \right) \mathfrak S_5.
\end{align}

By the same token as the previous case, one has for D(3-1)-brane the differential equations derived from the second scalar constraint, the $(rr)$ component of Einstein equation and the EOM of the scalar $A$ as
\begin{align}\label{eq: D(3-1) 2nd order scalar constraint 2}
  & 3(r^4 k_{(2)})' + 24 r^3 j_{(2)} - 2(3r^4-1) h_{(2)}' = \left( -4r + 3 r^2 F + r^3 F' \right) \mathbf s_2 \cr
  & + \left( 2r + 3 r^2 F + r^3 F' \right) \mathfrak S_1 + \bigg[ - \frac32 r + r^2 F + \frac83 r^3 F' + \frac13 r^4 F'' - \frac12 (3r^4-1) FF' \cr
  & + \frac18 r^5 f F'^2 - 6 r^2 F_j + 36 r^3 F_j^2 - 2(3r^4-1) F' F_j - 3 r^4 F' F_k - \frac12 r^5 F' F_k' + 24 r^3 F_j F_k \cr
  & + 6 r^4 F_j F_k' \bigg] \mathfrak S_3 + \left( \frac1{r^3} + 6r \right) \mathfrak S_4 - \left[ 5r + (3r^4-1) FF' + \frac14 r^5 f F'^2 \right] \mathfrak S_5, \\
  & r h_{(2)}'' + 2 h_{(2)}' - 3 j_{(2)}' = \left( \frac12 FF' + \frac14 r FF'' + \frac18 r F'^2 + \frac12 r F' F_j' - 3 F_j F_j' \right) \mathfrak S_3 \cr
  & \hspace{10em} + \frac1{r^3} \mathfrak S_4 + \left( FF' + \frac12 r FF'' + \frac14 r F'^2 \right) \mathfrak S_5, \\
  & (r^4 k_{(2)})' + r^4 f j_{(2)}' + 8 r^3 j_{(2)} - r^4 f h_{(2)}' = \left( -r + \frac12 r^2 F \right) \mathbf s_2 + \left( r + \frac12 r^2 F \right) \mathfrak S_1   \cr
  & + \bigg[ - \frac12 r + \frac16 r^2 F + \frac16 r^3 F' - \frac14 r^4 f FF' - r^2 F_j + 12 r^3 F_j^2 + 3 r^4 f F_j F_j' - r^4 f F' F_j  \cr
  & - \frac12 r^4 F' F_k + 8 r^3 F_j F_k + r^4 F_j' F_k + 2 r^4 F_j F_k'  \bigg] \mathfrak S_3 + r \mathfrak S_4 - \left( r + \frac12 r^4 f FF' \right) \mathfrak S_5. \label{eq: D(3-1) 2nd order EOM A}
\end{align}
The solutions of the above are
\begin{align}
  h ^{(2)} =& \left( \frac14 - \frac{\ln2}{8} \right) \frac{1}{r^4} ( \mathbf s_2 + \mathfrak S_1 ) + \left[ \frac{1}{4r^2} + \left( \frac1{24} - \frac{\ln2}{24} \right)\frac{1}{r^4} \right] \mathfrak S_3 \cr
  & + \frac1{2r^2} \mathfrak S_4 + \left( \frac1{2 r^2} + \frac{1}{4 r^4} \right) \mathfrak S_5, \\
  j^{(2)} =& - \left( \frac14 - \frac{\ln2}{8} \right) \frac{1}{r^4} ( \mathbf s_2 + \mathfrak S_1 ) - \left( \frac1{24} - \frac{\ln2}{24} \right) \frac{1}{r^4} \mathfrak S_3 - \frac1{4 r^4} \mathfrak S_5, \\
  k^{(2)} =& \left[ - \left( \frac16 - \frac{\ln2}{12} \right) \frac{1}{r^4} + \frac{1}{18 r^6} - \left( \frac14 - \frac{\ln2}{8} \right)\frac{1}{r^8} \right] \mathbf s_2 \cr
  & + \left[ \frac{1}{r^2} - \left( \frac16 - \frac{\ln2}{12} \right) \frac{1}{r^4} + \frac{1}{18 r^6} - \left( \frac14 - \frac{\ln2}{8} \right) \frac{1}{r^8} \right] \mathfrak S_1 \cr
  & + \left[ - \left( \frac1{36} - \frac{\ln2}{36} \right) \frac{1}{r^4} + \frac{1}{15 r^5} + \frac{37}{108 r^6} + \frac{\ln2}{24 r^8} \right] \mathfrak S_3 \cr
  & - \frac1{2 r^6} \mathfrak S_4 - \left( \frac1{6 r^4} - \frac2{5 r^5} - \frac1{6 r^6} + \frac1{4 r^8} \right) \mathfrak S_5.
\end{align}

For the situations of one spatial dimension after compactification, i.e. $p-q=1$, the D(4-3)-brane has the differential equations as
\begin{align}
  & 5(r^3 k_{(2)})' + 30 r^2 j_{(2)} - (5r^3-2) h_{(2)}' = \left( -16r + 5 r^\frac32 F + 2 r^\frac52 F' \right) \mathbf s_2 \cr
  & + \left( 24r + 5r^\frac32 F + 2r^\frac52 F' \right) \mathfrak S_1 + \bigg[ - 16r + r^\frac32 F + \frac{26}{5} r^\frac52 F' + \frac45 r^\frac72 F'' - (5r^3-2) FF' \cr
  & - 10 r^\frac32 F_j + 45 r^2 F_j^2 - 2(5r^3-2) F'F_j - 5r^3 F'F_k - r^4 F' F_k' + 30 r^2 F_j F_k \cr
  & + 10 r^3 F_j F_k' \bigg] \mathfrak S_3, \label{eq: D(4-3) 2nd order scalar constraint 2} \\
  & 2r h_{(2)}'' + 3 h_{(2)}' - 10 j_{(2)}' = \left( 3 FF' + 2 r FF'' + r F'^2 + 2r F' F_j' - 10 F_j F_j' \right) \mathfrak S_3, \\
  & (r^3 k_{(2)})' + r^3 f j_{(2)}' + 6 r^2 j_{(2)} - \frac12 r^3 f h_{(2)}' = \left( -2r + \frac12 r^\frac32 F \right) \mathbf s_2 + \left( 6r + \frac12 r^\frac32 F \right) \mathfrak S_1   \cr
  & + \bigg[ - 2r + \frac{1}{10} r^\frac32 F + \frac15 r^\frac52 F' - \frac12 r^3 f FF' - r^\frac32 F_j + 9 r^2 F_j^2 + 3 r^3 f F_j F_j' - r^3 f F' F_j  \cr
  & - \frac12 r^3 F' F_k + 6 r^2 F_j F_k + r^3 F_j' F_k + 2 r^3 F_j F_k'  \bigg] \mathfrak S_3.
\end{align}
Which can be solved as
\begin{align}
  h ^{(2)} =& \left( \frac43 - \frac{\pi}{9\sqrt3} - \frac{\ln3}{3} \right) \frac{1}{r^3} ( \mathbf s_2 + \mathfrak S_1 ) + \left[ \frac{4}{r} + \left( \frac{16}{15} - \frac{\pi}{45\sqrt3} - \frac{\ln3}{15} \right)\frac{1}{r^3} \right] \mathfrak S_3, \\
  j^{(2)} =& - \left( \frac23 - \frac{\pi}{18\sqrt3} - \frac{\ln3}{6} \right) \frac{1}{r^3} ( \mathbf s_2 + \mathfrak S_1 ) - \left( \frac{5}{18} - \frac{\pi}{90\sqrt3} - \frac{\ln3}{30} \right) \frac{1}{r^3} \mathfrak S_3, \\
  k^{(2)} =& \left[ - \left( \frac{4}{15} - \frac{\pi}{45\sqrt3} - \frac{\ln3}{15} \right) \frac{1}{r^3} + \frac{1}{10 r^4} - \left( \frac23 - \frac{\pi}{18\sqrt3} - \frac{\ln3}{6} \right)\frac{1}{r^6} \right] \mathbf s_2 \cr
  & + \left[ \frac{4}{r} - \left( \frac{4}{15} - \frac{\pi}{45\sqrt3} - \frac{\ln3}{15} \right) \frac{1}{r^3} + \frac{1}{10 r^4} - \left( \frac23 - \frac{\pi}{18\sqrt3} - \frac{\ln3}{6} \right) \frac{1}{r^6} \right] \mathfrak S_1 \cr
  & + \left[ - \left( \frac{16}{75} - \frac{\pi}{225\sqrt3} - \frac{\ln3}{75} \right) \frac{1}{r^3} + \frac{32}{35 r^{7/2}} + \frac{119}{50 r^4} \right. \cr
  & \left. - \left( \frac{22}{45} - \frac{\pi}{90\sqrt3} - \frac{\ln3}{30} \right) \frac{1}{r^6} \right] \mathfrak S_3.
\end{align}

For the D(3-2)-brane, we have
\begin{align}
  & 3(r^4 k_{(2)})' + 24 r^3 j_{(2)} - (3r^4-1) h_{(2)}' = \left( -4r + 3 r^2 F + r^3 F' \right) \mathbf s_2 \cr
  & + \left( 2r + 3r^2 F + r^3 F' \right) \mathfrak S_1 + \bigg[ - 4r + r^2 F + \frac83 r^3 F' + \frac13 r^4 F'' - (3r^4-1) FF' \cr
  & - 6 r^2 F_j + 36 r^3 F_j^2 - 2(3r^4-1) F' F_j - 3r^4 F' F_k - \frac12 r^5 F' F_k' + 24 r^3 F_j F_k \cr
  & + 6 r^4 F_j F_k' \bigg] \mathfrak S_3, \label{eq: D(3-2) 2nd order scalar constraint 2} \\
  & r h_{(2)}'' + 2 h_{(2)}' - 6 j_{(2)}' = \left( 2 FF' + r FF'' + \frac12 r F'^2 - 6 F_j F_j' + r F' F_j' \right) \mathfrak S_3, \\
  & (r^4 k_{(2)})' + r^4 f j_{(2)}' + 8 r^3 j_{(2)} - \frac12 r^4 f h_{(2)}' = \left( -r + \frac12 r^2 F \right) \mathbf s_2 + \left( r + \frac12 r^2 F \right) \mathfrak S_1   \cr
  & + \bigg[ - r + \frac16 r^2 F + \frac16 r^3 F' - \frac12 r^4 f FF' - r^2 F_j + 12 r^3 F_j^2 + 3 r^4 f F_j F_j' - r^4 f F' F_j  \cr
  & - \frac12 r^4 F' F_k + 8 r^3 F_j F_k + r^4 F_j' F_k + 2 r^4 F_j F_k'  \bigg] \mathfrak S_3.
\end{align}
Note the last differential equation is from the EOM of $A$. The solutions of the above are
\begin{align}
  h ^{(2)} =& \left( \frac12 - \frac{\ln2}{4} \right) \frac{1}{r^4} ( \mathbf s_2 + \mathfrak S_1 ) + \left[ \frac{1}{r^2} + \left( \frac13 - \frac{\ln2}{12} \right)\frac{1}{r^4} \right] \mathfrak S_3, \\
  j^{(2)} =& - \left( \frac14 - \frac{\ln2}{8} \right) \frac{1}{r^4} ( \mathbf s_2 + \mathfrak S_1 ) - \left( \frac16 - \frac{\ln2}{24} \right) \frac{1}{r^4} \mathfrak S_3, \\
  k^{(2)} =& \left[ - \left( \frac16 - \frac{\ln2}{12} \right) \frac{1}{r^4} + \frac{1}{18 r^6} - \left( \frac14 - \frac{\ln2}{8} \right)\frac{1}{r^8} \right] \mathbf s_2 \cr
  & + \left[ \frac{1}{r^2} - \left( \frac16 - \frac{\ln2}{12} \right) \frac{1}{r^4} + \frac{1}{18 r^6} - \left( \frac14 - \frac{\ln2}{8} \right) \frac{1}{r^8} \right] \mathfrak S_1 \cr
  & + \left[ - \left( \frac19 - \frac{\ln2}{36} \right) \frac{1}{r^4} + \frac{4}{15 r^5} + \frac{23}{54 r^6} - \left( \frac18 - \frac{\ln2}{24} \right) \frac1{r^8} \right] \mathfrak S_3.
\end{align}

The differential equations for the second order scalar perturbations of D(2-1)-brane can be got as
\begin{align}
  & 7(r^5 k_{(2)})' + 70 r^4 j_{(2)} - (7r^5-2) h_{(2)}' = \left( -\frac{16}{3} r + 7 r^\frac52 F + 2 r^\frac72 F' \right) \mathbf s_2 \cr
  & + \left( \frac89 r + 7 r^\frac52 F + 2 r^\frac72 F' \right) \mathfrak S_1 + \bigg[ - \frac{16}3 r + 3 r^\frac52 F + \frac{38}{7} r^\frac72 F' + \frac47 r^\frac92 F'' - (7r^5-2) FF' \cr
  & - 14 r^\frac52 F_j + 105 r^4 F_j^2 - 2(7r^5-2) F' F_j - 7r^5 F' F_k - r^6 F' F_k' + 70 r^4 F_j F_k \cr
  & + 14 r^5 F_j F_k' \bigg] \mathfrak S_3, \label{eq: D(2-1) 2nd order scalar constraint 2} \\
  & 2r h_{(2)}'' + 5 h_{(2)}' - 14 j_{(2)}' = \left( 5 FF' + 2 r FF'' + r F'^2 - 14 F_j F_j' + 2r F' F_j' \right) \mathfrak S_3, \\
  & (r^5 k_{(2)})' + r^5 f j_{(2)}' + 10 r^4 j_{(2)} - \frac12 r^5 f h_{(2)}' = \left( -\frac23 r + \frac12 r^\frac52 F \right) \mathbf s_2 + \left( \frac29 r + \frac12 r^\frac52 F \right) \mathfrak S_1   \cr
  & + \bigg[ - \frac23 r + \frac{3}{14} r^\frac52 F + \frac17 r^\frac72 F' - \frac12 r^5 f FF' - r^\frac52 F_j + 15 r^4 F_j^2 + 3 r^5 f F_j F_j' - r^5 f F' F_j  \cr
  & - \frac12 r^5 F' F_k + 10 r^4 F_j F_k + r^5 F_j' F_k + 2 r^5 F_j F_k'  \bigg] \mathfrak S_3.
\end{align}
Solving these perturbations is tough because $F(r)$ of D2-brane is more complex than the other Dp-brane, as can be seen from \cref{eq: Dp-brane F(r)}. One can simplify the calculation by first making the series expansion with respect to $1/r$ and then do the integral when solving the differential equations of $j^{(2)}$ and $k^{(2)}$. The solutions are
\begin{align}
  h ^{(2)} =& \left( \frac{4}{15} + \frac{\pi}{25} \sqrt{1 - \frac2{\sqrt5}} + \frac{1}{5\sqrt5} \,{\rm arcoth}\,\sqrt5 - \frac{\ln5}{10} \right) \frac{1}{r^5} ( \mathbf s_2 + \mathfrak S_1 ) \cr
  & + \left[ \frac{4}{9r^3} + \left( \frac{16}{105} + \frac{3\pi}{175} \sqrt{1 - \frac2{\sqrt5}} + \frac{3}{35\sqrt5} \,{\rm arcoth}\,\sqrt5 - \frac{3}{70} \ln5 \right) \frac{1}{r^5} \right] \mathfrak S_3, \\
  j^{(2)} =& - \left( \frac{2}{15} + \frac{\pi}{50} \sqrt{1 - \frac2{\sqrt5}} + \frac{1}{10\sqrt5} \,{\rm arcoth}\,\sqrt5 - \frac{\ln5}{20} \right) \frac{1}{r^5} ( \mathbf s_2 + \mathfrak S_1 ) \cr
  & - \left( \frac{8}{105} + \frac{3\pi}{350} \sqrt{1 - \frac2{\sqrt5}} + \frac{3}{70\sqrt5} \,{\rm arcoth}\,\sqrt5 - \frac{3\ln5}{140} \right) \frac{1}{r^5} \mathfrak S_3, \\
  k^{(2)} =& \bigg[ - \left( \frac{4}{35} + \frac{3\pi}{175} \sqrt{1 - \frac2{\sqrt5}} + \frac{3}{35\sqrt5} \,{\rm arcoth}\,\sqrt5 - \frac{3\ln5}{70} \right) \frac{1}{r^5} + \frac{1}{28 r^8} \cr
  & - \left( \frac{2}{15} + \frac{\pi}{50} \sqrt{1 - \frac2{\sqrt5}} + \frac{1}{10\sqrt5} \,{\rm arcoth}\,\sqrt5 - \frac{\ln5}{20} \right) \frac{1}{r^{10}} \bigg] \mathbf s_2 \cr
  & + \bigg[ \frac4{9r^3} - \left( \frac{4}{35} + \frac{3\pi}{175} \sqrt{1 - \frac2{\sqrt5}} + \frac{3}{35\sqrt5} \,{\rm arcoth}\,\sqrt5 - \frac{3\ln5}{70} \right) \frac{1}{r^5} + \frac{1}{28r^8} \cr
  & - \left( \frac{2}{15} + \frac{\pi}{50} \sqrt{1 - \frac2{\sqrt5}} + \frac{1}{10\sqrt5} \,{\rm arcoth}\,\sqrt5 - \frac{\ln5}{20} \right) \frac{1}{r^{10}} \bigg] \mathfrak S_1 \cr
  & + \bigg[ - \left( \frac{16}{245} + \frac{9\pi}{1225} \sqrt{1 - \frac2{\sqrt5}} + \frac{9}{245\sqrt5} \,{\rm arcoth}\,\sqrt5 - \frac{9 \ln5}{490} \right) \frac{1}{r^5} + \frac{32}{ 273 r^\frac{13}{2} } \cr
  & + \frac{253}{1764 r^8} - \left( \frac{22}{525} + \frac{3\pi}{350} \sqrt{1 - \frac2{\sqrt5}} + \frac{3}{70\sqrt5} \,{\rm arcoth}\,\sqrt5 - \frac{3 \ln5}{140} \right) \frac{1}{r^{10}} \bigg] \mathfrak S_3.
\end{align}

\section{The second order constitutive relation} \label{sec: constitutive relations of the 2nd order}

In this section, we will offer all the results of the second order constitutive relations for the D(p-q)-brane with $2\leq p\leq 4$ and $1\leq q\leq p-1$. We will omit D(4-1)-brane case here as in the previous sections, for its results can be found in \cite{Wu1604}. One will find the following substitutions very helpful to gain the final form of the constitutive relations:
\begin{align}
  &t_{3ij}=\partial_0\sigma_{ij} \to \sideset{_\langle}{}{\mathop D}\partial_\mu u_{\nu\rangle}, \qquad \mathfrak T_{1ij} = \partial_0\beta_i \partial_0\beta_j - \frac1{p-q} \delta_{ij}\mathbf s_1 \to Du_{\langle\mu}Du_{\nu\rangle}, \cr
  &\mathfrak T_{4ij} = \sigma_{ij}\partial\beta \to \sigma_{\mu\nu} \partial_\rho u^\rho, \qquad \mathfrak T_{7ij} = \sigma_{(i}^{~~k} \Omega_{j)k} \to \sigma_{\langle\mu}^{~~\rho} \Omega_{\nu\rangle\rho}
\end{align}
are for the spatial viscous tensors and
\begin{align}
  \mathbf s_2 + \mathfrak S_1 = \partial_0 \partial \beta + \partial_0\beta_i \partial_0\beta_i \to D\partial u, \quad \mathfrak S_3 = (\partial \beta)^2 \to (\partial u)^2, \quad \mathfrak S_5 = \sigma_{ij}^2 \to \sigma_{\mu\nu}^2
\end{align}
are for the spatial viscous scalar terms. Here the angle brackets stand for the transverse traceless symmetrization of rank 2 tensors, i.e.
\begin{align}
  A_{\langle\mu\nu\rangle} = P_\mu^\rho P_\nu^\lambda A_{(\rho\lambda)} - \frac1{p-q} P_{\mu\nu} P^{\rho\lambda} A_{\rho\lambda}.
\end{align}

The situation of $p-q=2$ includes the D(4-2) and D(3-1)-brane cases. For the D(4-2)-brane, one has
\begin{align}\label{eq: D(4-2) 2nd order stress energy tensor}
  T_{\mu\nu} &= \frac1{2\kappa_4^2} \Bigg\{ \frac{r_H^3}{L_4^4} \bigg( \frac52 u_\mu u_\nu + \frac12 P_{\mu\nu} \bigg) - \left( \frac{r_H}{L_4} \right)^\frac52 \bigg( 2\sigma_{\mu\nu} + \frac35 \partial_\rho u^\rho P_{\mu\nu} \bigg) \cr
  & + \frac{r_H^2}{L_4} \bigg[ \bigg( 2 - \frac\pi{6\sqrt3} - \frac{\ln3}2 \bigg) \cdot 2\bigg( \sideset{_\langle}{}{\mathop D}\sigma_{\mu\nu\rangle} + \frac12 \sigma_{\mu\nu} \partial u \bigg) + \bigg( \frac65 + \frac{\pi}{10\sqrt3} + \frac{3\ln3}{10} \bigg) \frac{ 2\sigma_{\mu\nu}\partial u}{2} \cr
  & - \bigg( \frac{\pi}{3\sqrt3} + \ln3\bigg ) \cdot 2\sigma_{\langle\mu}^{~~\rho} \Omega_{\nu\rangle\rho} \bigg] + \frac{r_H^2}{L_4} P_{\mu\nu} \bigg[ \bigg( \frac65 - \frac{\pi}{10\sqrt3} - \frac{3\ln3}{10} \bigg) D(\partial u) \cr
  & + \bigg( \frac9{25} - \frac{\pi}{50\sqrt3} - \frac{3\ln3}{50} \bigg) (\partial u)^2 + \frac3{10}\cdot 4\sigma_{\rho \lambda}^2 \bigg] \Bigg\}.
\end{align}
With the second order transport coefficients can be read from the above as
\begin{align}
  \eta\tau_\pi = \frac1{2\kappa_4^2} \bigg( 2 - \frac\pi{6\sqrt3} - \frac{\ln3}2 \bigg) \frac{r_H^2}L, &\qquad \eta\tau_\pi^* = \frac1{2\kappa_4^2} \bigg( \frac65 + \frac{\pi}{10\sqrt3} + \frac{3\ln3}{10} \bigg) \frac{r_H^2}L, \cr
  \lambda_2 = - \frac1{2\kappa_4^2} \bigg( \frac{\pi}{3\sqrt3} + \ln3\bigg ) \frac{r_H^2}L, &\qquad \zeta\tau_\Pi = \frac1{2\kappa_4^2} \bigg( \frac65 - \frac{\pi}{10\sqrt3} - \frac{3\ln3}{10} \bigg) \frac{r_H^2}L, \cr
  \xi_1 = \frac1{2\kappa_4^2} \frac{3}{10} \frac{r_H^2}L, &\qquad \xi_2 = \frac1{2\kappa_4^2} \bigg( \frac9{25} - \frac{\pi}{50\sqrt3} - \frac{3\ln3}{50} \bigg) \frac{r_H^2}L.
\end{align}

The D(3-1)-brane has the second order stress energy tensor as
\begin{align}\label{}
  T_{\mu\nu} &= \frac1{2\kappa_4^2} \Bigg\{ \frac{r_H^4}{L_3^5} \bigg( 3 u_\mu u_\nu + P_{\mu\nu} \bigg) - \left( \frac{r_H}{L_3} \right)^3 \bigg( 2\sigma_{\mu\nu} + \frac13 P_{\mu\nu} \partial_\rho u^\rho \bigg) \cr
  & + \frac{r_H^2}{L_3} \bigg[ \left( 1 - \frac{\ln2}{2} \right) \cdot 2\bigg( \sideset{_\langle}{}{\mathop D}\sigma_{\mu\nu\rangle} + \frac12 \sigma_{\mu\nu} \partial u \bigg) + \left( \frac13 + \frac{\ln2}{6} \right) \frac{ 2\sigma_{\mu\nu}\partial u}{2} - \ln2 \cdot 2\sigma_{\langle\mu}^{~~\rho} \Omega_{\nu\rangle\rho} \bigg] \cr
  & + P_{\mu\nu} \frac{r_H^2}{L_3} \bigg[ \left( \frac13 - \frac{\ln2}{6} \right) D(\partial u) + \left( \frac{1}{18} - \frac{\ln2}{18} \right) (\partial u)^2 + \frac1{12} \cdot 4\sigma_{\rho \lambda}^2 \bigg] \Bigg\}.
\end{align}
The corresponding second order transport coefficients are
\begin{align}
  \eta\tau_\pi &= \frac1{2\kappa_4^2} \left( 1 - \frac{\ln2}{2} \right) \frac{r_H^2}{L_3}, \quad \eta\tau_\pi^* = \frac1{2\kappa_4^2} \left( \frac13 + \frac{\ln2}{6} \right) \frac{r_H^2}{L_3},\quad\lambda_2 = - \frac1{2\kappa_4^2} \ln2 \frac{r_H^2}{L_3}, \cr
  \zeta\tau_\Pi &= \frac1{2\kappa_4^2} \left( \frac13 - \frac{\ln2}{6} \right) \frac{r_H^2}{L_3},\quad\xi_1 = \frac1{2\kappa_4^2} \frac1{12} \frac{r_H^2}{L_3},\qquad\xi_2 = \frac1{2\kappa_4^2} \left( \frac{1}{18} - \frac{\ln2}{18} \right) \frac{r_H^2}{L_3}.
\end{align}

When $p-q=1$, one has the D(4-3), D(3-2) and D(2-1)-brane. These cases are like the D1-brane in \cite{Wu1807}. The spatial viscous terms do not contain $\sigma_{ij}$ or $\Omega_{ij}$ since there is only one spatial direction. Thus the results will be much simpler than the cases of $p-q=2$. The second order constitutive relation for D(4-3)-brane is
\begin{align}\label{}
  T_{\mu\nu} &= \frac1{2\kappa_3^2} \Bigg\{ \frac{r_H^3}{L_4^4} \bigg( \frac52 u_\mu u_\nu + \frac12 P_{\mu\nu} \bigg) - \left( \frac{r_H}{L_4} \right)^\frac52 \frac85 P_{\mu\nu} \partial_\rho u^\rho \cr
  & + \frac{r_H^2}{L_4} P_{\mu\nu} \bigg[ \bigg( \frac{16}5 - \frac{4\pi}{15\sqrt3} - \frac{4\ln3}{5} \bigg) D(\partial u) + \bigg( \frac{64}{25} - \frac{4\pi}{75\sqrt3} - \frac{4\ln3}{25} \bigg) (\partial u)^2 \bigg] \Bigg\}. \quad
\end{align}
There are only two transport coefficients at the second order:
\begin{align}
  \zeta\tau_\Pi = \frac1{2\kappa_3^2} \bigg( \frac{16}5 - \frac{4\pi}{15\sqrt3} - \frac{4\ln3}{5} \bigg) \frac{r_H^2}{L_4}, \qquad
  \xi_2 = \frac1{2\kappa_3^2} \bigg( \frac{64}{25} - \frac{4\pi}{75\sqrt3} - \frac{4\ln3}{25} \bigg) \frac{r_H^2}{L_4}
\end{align}
In the D(3-2)-brane case one has
\begin{align}\label{}
  T_{\mu\nu} =&\; \frac1{2\kappa_3^2} \Bigg\{ \frac{r_H^4}{L_3^5} \bigg( 3 u_\mu u_\nu + P_{\mu\nu} \bigg) - \left( \frac{r_H}{L_3} \right)^3 \frac43 P_{\mu\nu} \partial_\rho u^\rho \cr
  & + P_{\mu\nu} \frac{r_H^2}{L_3} \left[ \left( \frac43 - \frac{2\ln2}{3} \right) D(\partial u) + \left( \frac89 - \frac{2\ln2}{9} \right) (\partial u)^2 \right] \Bigg\},
\end{align}
whose second order transport coefficients can be read as
\begin{align}
  \zeta\tau_\Pi = \frac1{2\kappa_3^2} \left( \frac43 - \frac{2\ln2}{3} \right) \frac{r_H^2}{L_3}, \qquad
  \xi_2 = \frac1{2\kappa_3^2} \left( \frac89 - \frac{2\ln2}{9} \right) \frac{r_H^2}{L_3}.
\end{align}
The situation of D(2-1)-brane gives us
\begin{align}
  T_{\mu\nu} &= \frac{1}{2\kappa_3^2} \left\{ \frac{r_H^5}{L_2^6} \left( \frac72 u_\mu u_\nu + \frac32 P_{\mu\nu} \right) - \left( \frac{r_H}{L_2} \right)^\frac72 \frac87 P_{\mu\nu} \partial u \right. \cr
  & + P_{\mu\nu} \frac{r_H^2}{L_2} \left[ \left( \frac{16}{21} + \frac{4\pi}{35} \sqrt{1 - \frac2{\sqrt5}} + \frac{4}{7\sqrt5} \,{\rm arcoth}\,\sqrt5 - \frac{2\ln5}{7} \right) D\partial u \right. \cr
  & + \left. \left. \left( \frac{64}{147} + \frac{12\pi}{245} \sqrt{1 - \frac2{\sqrt5}} + \frac{12}{49\sqrt5} \,{\rm arcoth}\,\sqrt5 - \frac{6\ln5}{49} \right) (\partial u)^2 \right] \right\},
\end{align}
with the second order coefficients are
\begin{align}
  \zeta\tau_\Pi &= \frac1{2\kappa_3^2} \left( \frac{16}{21} + \frac{4\pi}{35} \sqrt{1 - \frac2{\sqrt5}} + \frac{4}{7\sqrt5} \,{\rm arcoth}\,\sqrt5 - \frac{2\ln5}{7} \right) \frac{r_H^2}{L_2}, \cr
   \xi_2 &= \frac1{2\kappa_3^2} \left( \frac{64}{147} + \frac{12\pi}{245} \sqrt{1 - \frac2{\sqrt5}} + \frac{12}{49\sqrt5} \,{\rm arcoth}\,\sqrt5 - \frac{6\ln5}{49} \right) \frac{r_H^2}{L_2}.
\end{align}

All the results listed in this section can be rewritten in a universal form by introducing the Harmonic number and the second order constitutive relations for the Dp-brane with $q$ directions of their world-volume compactified can be recast in the form as
\begin{align}\label{eq: D(p-q) 2nd order stress-energy tensor (general)}
  T_{\mu\nu} =&\; \frac{1}{2 \kappa_{p-q+2}^2} \Bigg\{ {r_H^{7-p} \over L_p^{8-p}} \left( \frac{9-p}{2} u_\mu u_\nu + \frac{5-p}{2} P_{\mu\nu} \right) \cr
  & - \left( \frac{r_H}{L_p} \right) ^\frac{9-p}{2} \bigg( 2\sigma_{\mu\nu} + \frac{2(p-3)^2 + 2q(5-p)}{(p-q)(9-p)} P_{\mu\nu} \partial u \bigg) \cr
  & + \frac{r_H^2}{L_p} \Bigg[ \bigg( \frac{1}{5-p} + \frac{1}{7-p} H_\frac{5-p}{7-p} \bigg)\cdot 2\bigg( \sideset{_\langle}{}{\mathop D}\sigma_{\mu\nu\rangle} + \frac1{p-q} \sigma_{\mu\nu} \partial u \bigg) \cr
  & + \bigg( \frac{3(p-3)^2 + 3q(5-p)}{(5-p)(9-p)} - \frac{(p-3)^2 + q(5-p)}{(7-p)(9-p)} H_\frac{5-p}{7-p} \bigg) \frac{2\sigma_{\mu\nu} \partial u}{p-q} \cr
  & + \frac{1}{5-p} \cdot 4\sigma_{\langle\mu}^{~~\rho}\sigma_{\nu\rangle\rho} + \bigg( - \frac{2}{5-p} + \frac{2}{7-p} H_\frac{5-p}{7-p} \bigg) \cdot 2 \sigma_{\langle\mu}^{~~\rho} \Omega_{\nu\rangle\rho} \Bigg] \cr
  & + P_{\mu\nu} \frac{r_H^2}{L_p} \Bigg[ \bigg( \frac{2(p-3)^2 + 2q(5-p)}{(p-q)(5-p)(9-p)} + \frac{2(p-3)^2 + 2q(5-p)}{(p-q)(7-p)(9-p)} H_\frac{5-p}{7-p} \bigg) D(\partial u) \cr
  & + \Bigg( \frac{[2(p-3)^2 + 2q(5-p)] [(3p^2 - 17p + 18) + 3q(5-p)]}{(p-q)^2 (5-p)(9-p)^2} \cr
  & + \frac{(5-p) [2(p-3)^2 + 2q(5-p)]}{(p-q) (7-p) (9-p)^2} H_\frac{5-p}{7-p} \Bigg) (\partial u)^2 \cr
  & + \frac{(p-3)^2 + q(5-p)}{(p-q)(5-p)(9-p)} \cdot 4 \sigma_{\alpha\beta}^2 \Bigg] \Bigg\}.
\end{align}
from which one can read all the second order transport coefficients as
\begin{align}\label{eq: D(p-q) 2nd order coefficients}
  \eta\tau_\pi &= \frac1{2\kappa_{p-q+2}^2} \left( \frac{1}{5-p} + \frac{1}{7-p} H_\frac{5-p}{7-p} \right) \frac{r_H^2}{L_p}, \cr
  \eta\tau_\pi^* &= \frac1{2\kappa_{p-q+2}^2} \left[ \frac{3(p-3)^2 + 3q(5-p)}{(5-p)(9-p)} - \frac{(p-3)^2 + q(5-p)}{(7-p)(9-p)} H_\frac{5-p}{7-p} \right] \frac{r_H^2}{L_p}, \cr
  \lambda_1 &= \frac1{2\kappa_{p-q+2}^2} \frac{1}{5-p} \frac{r_H^2}{L_p}, \qquad \lambda_2 = \frac1{2\kappa_{p-q+2}^2} \left( - \frac{2}{5-p} + \frac{2}{7-p} H_\frac{5-p}{7-p} \right) \frac{r_H^2}{L_p}, \cr
  \zeta\tau_\Pi &= \frac1{2\kappa_{p-q+2}^2} \Bigg[ \frac{2(p-3)^2 + 2q(5-p)}{(p-q)(5-p)(9-p)} + \frac{2(p-3)^2 + 2q(5-p)}{(p-q)(7-p)(9-p)} H_\frac{5-p}{7-p} \Bigg] \frac{r_H^2}{L_p}, \cr
  \xi_1 &= \frac1{2\kappa_{p-q+2}^2} \frac{(p-3)^2 + q(5-p)}{(p-q)(5-p)(9-p)} \frac{r_H^2}{L_p}, \cr
  \xi_2 &= \frac1{2\kappa_{p-q+2}^2} \Bigg[ \frac{[2(p-3)^2 + 2q(5-p)] [(3p^2 - 17p + 18) + 3q(5-p)]}{(p-q)^2 (5-p)(9-p)^2} \cr
  & \quad + \frac{(5-p) [2(p-3)^2 + 2q(5-p)]}{(p-q) (7-p) (9-p)^2} H_\frac{5-p}{7-p} \Bigg] \frac{r_H^2}{L_p}.
\end{align}
We have added the result of D(4-1)-brane into the above formulas, that's why the viscous tensor relates with $\lambda_1$ appears. The definition of the Harmonic number $H_\frac{5-p}{7-p}$ with its special value for the cases of Dp-brane with $1\leq p\leq 4$ can be found in \cite{Wu1807}. \cref{eq: D(p-q) 2nd order stress-energy tensor (general),eq: D(p-q) 2nd order coefficients} will reproduce the results in \cite{Wu1807} and \cite{Wu1604} by separately setting $q=0$ and $p=4,~q=1$.

One can easily see that of the 7 second order transport coefficients that we derived, only $\eta\tau_\pi^*,\ \zeta\tau_\Pi,\ \xi_1, \ \xi_2$ depend on the number of the compact directions of the brane, i.e., their expressions contain $q$  while $\eta\tau_\pi,\ \lambda_1,\ \lambda_2$ do not. If we take a closer look at how $q$ enters the formulae, we will find that $q$ appears in the form of
\begin{align}
  \frac{2(p-3)^2 + 2q(5-p)}{(p-q)(9-p)},
\end{align}
which is just the numerical factor of bulk viscosity $\zeta$. Only $\eta \tau_\pi^*$ is a little special. The $p-q$ in the denominator is with the viscous tensor $\frac{2\sigma_{\mu\nu} \partial u}{p-q}$ by definition.

Since the number of spatial dimensions of the compactified Dp-brane is also the number of the spatial dimensions of the boundary fluid. Compactifying the Dp-brane will reduce the number of the spatial dimensions of the boundary fluid, too.
For the boundary fluid, the bulk flow viscous terms $\partial u,\ D(\partial u),\ (\partial u)^2$ and $\sigma_{\mu\nu} \partial u$ contain the sum of diagonal spatial components and the bulk flow viscous term $\sigma_{\alpha\beta}^2$ sums the nondiagonal spatial components. There will always be some spatial components lost if one or more spatial directions of Dp-branes are compactified. So the bulk flow viscous terms are affected by compactification and their transport coefficients $\eta\tau_\pi^*,\ \zeta\tau_\Pi,\ \xi_1$ and $\xi_2$ should include $q$.
On the contrary,  the shear flow viscous terms $\sideset{_\langle}{}{\mathop D}\sigma_{\mu\nu\rangle},\ \sigma_{\langle\mu}^{~~\rho}\sigma_{\nu\rangle\rho},\ \sigma_{\langle\mu}^{~~\rho} \Omega_{\nu\rangle\rho}$ are tensorial in nature. By definition they do not sum spatial components. Thus their transport coefficients $\eta\tau_\pi,~\lambda_1,~\lambda_2$ should not be under affected by compactification and do not contain $q$.

As one can check, the results of \cref{eq: D(p-q) 2nd order coefficients} satisfy the Haack-Yarom relation\footnote{It is firstly found in 5 dimensional charged AdS black hole \cite{Erdmenger0809} and is firstly confirmed in nonconformal regime holographically in \cite{Bigazzi1006}.} $4\lambda_1 + \lambda_2 = 2 \eta \tau_\pi$, the Romatschke relations \cite{Romatschke0906}
\begin{align}
  \tau_\pi = \tau_\Pi, \qquad \xi_1 = \frac{1}{p-q} [1 - (p-q) c_s^2] \lambda_1
\end{align}
and also the Kleinert-Probst relations \cite{Kleinert1610}
\begin{align}\label{eq: Kleinert-Probst relation (generalized)}
  \eta\tau^*_\pi &= \left(1 - (p-q) c_s^2 \right) (4\lambda_1 - \eta\tau_\pi), \cr
  \xi_2 &= \frac2{(p-q)^2} \left(1 - (p-q) c_s^2 \right) \left[ (1 - 2(p-q) c_s^2) 2\lambda_1 + (p-q) c_s^2 \eta\tau_\pi \right].
\end{align}
Since $p-q$ is the spatial dimension of the dual relativistic fluid, if we denote $d$ as the dimension of the relativistic fluid, then we have $p-q=d-1$. So the above Romatschke and Kleinert-Probst relations still have the same form as in \cite{Wu1807}. The above 5 relations are among the 7 dynamical transport coefficients, one may draw a conclusion that only 2 of the 7 dynamical transport coefficients are actually independent. For which one can choose $\eta \tau_\pi$ and $\lambda_1$, since they appear in every relation. Ref. \cite{Bhattacharyya1201} derives 5 relations among the 8 thermodynamical second order transport coefficients\footnote{Following the nomenclature of \cite{Moore1210}.}, so only 3 of them are actually independent. For which one can choose $\lambda_3,\ \lambda_4$ and $\kappa$. Thus one may make a bold guess that at second order, nonconformal relativistic fluids have at most 5 independent flow modes:
\begin{align}
  \sideset{_\langle}{}{\mathop D} \sigma_{\mu\nu\rangle}, \quad \sigma_{\langle\mu}^{~~\rho}\sigma_{\nu\rangle\rho}, \quad \Omega_{\langle\mu}^{~~\rho}\Omega_{\nu\rangle\rho}, \quad
  \nabla_{\langle\mu}^\bot \ln s \nabla_{\nu\rangle}^\bot \ln s, \quad
  \left( R_{\langle\mu\nu\rangle} - 2 u^\rho u^\sigma R_{\rho\langle\mu\nu\rangle\sigma}\right),
\end{align}
which relate to the 5 independent second order transport coefficients. Here $\nabla_{\mu}^\bot = P_\mu^\rho \nabla_\rho$ is the covariant derivative perpendicular to the direction of $u^\mu$.

One can also reformulate the results of this work in terms of field theory quantities, which can be found in \cref{tab: reformulate the results in field theory language}. Compared with the field theory reformulation of the transport coefficients of Dp-brane \cite{Wu1807}, here we use $\Gamma\left( \frac{7-p}{2} \right)$ but not $\Gamma\left( \frac{9-p}{2} \right)$. Thus the dimensionless part of the results here is different from that of Dp-brane. We have also set the characteristic energy scale of the Dp-brane the same as the Hawking temperature, i.e. $\Lambda=T$ to make the expressions neat and clear. Another obvious difference from the Dp-brane case is that $V_q$ appears in all the thermal quantities as well as the transport coefficients which comes from the integration on the compact directions of Dp-branes.

\begin{table}
\renewcommand\arraystretch{2.}
\centering
\begin{tabular}{|c|c|c|}
\hline
 $\varepsilon$     & $\frac{9-p}{2} \left( \frac{4\pi}{7-p} \right)^2 { 2^\frac{2(6-p)}{5-p} \pi^\frac{3-p}{5-p} \over (7-p)^\frac{9-p}{5-p} } \Gamma\left( \frac{7-p}2 \right)^\frac{2}{5-p} \lambda_{p+1}^\frac{p-3}{5-p} N^2 V_q T^{p+1}$ \\[2mm] 
\hline
 $\mathfrak p$    & $\frac{5-p}{2} \left( \frac{4\pi}{7-p} \right)^2 { 2^\frac{2(6-p)}{5-p} \pi^\frac{3-p}{5-p} \over (7-p)^\frac{9-p}{5-p} } \Gamma\left( \frac{7-p}2 \right)^\frac{2}{5-p} \lambda_{p+1}^\frac{p-3}{5-p} N^2 V_q T^{p+1}$ \\[2mm]
\hline
 $\eta$                & $\left( \frac{4\pi}{7-p} \right) { 2^\frac{2(6-p)}{5-p} \pi^\frac{3-p}{5-p} \over (7-p)^\frac{9-p}{5-p} } \Gamma\left( \frac{7-p}2 \right)^\frac{2}{5-p} \lambda_{p+1}^\frac{p-3}{5-p} N^2 V_q T^p$ \\[2mm]
\hline
 $\zeta$              & $\frac{2(p-3)^2 + 2q(5-p)}{(p-q)(9-p)} \left( \frac{4\pi}{7-p} \right) { 2^\frac{2(6-p)}{5-p} \pi^\frac{3-p}{5-p} \over (7-p)^\frac{9-p}{5-p} } \Gamma\left( \frac{7-p}2 \right)^\frac{2}{5-p} \lambda_{p+1}^\frac{p-3}{5-p} N^2 V_q T^p$ \\[2mm]
\hline
 $\eta\tau_\pi$   & $\left( \frac{1}{5-p} + \frac{1}{7-p} H_\frac{5-p}{7-p} \right) { 2^\frac{2(6-p)}{5-p} \pi^\frac{3-p}{5-p} \over (7-p)^\frac{9-p}{5-p} } \Gamma\left( \frac{7-p}2 \right)^\frac{2}{5-p} \lambda_{p+1}^\frac{p-3}{5-p} N^2 V_q T^{p-1}$ \\[2mm]
\hline
 $\eta\tau_\pi^*$ & $\left[ \frac{3(p-3)^2 + 3q(5-p)}{(5-p)(9-p)} - \frac{(p-3)^2 + q(5-p)}{(7-p)(9-p)} H_\frac{5-p}{7-p} \right] { 2^\frac{2(6-p)}{5-p} \pi^\frac{3-p}{5-p} \over (7-p)^\frac{9-p}{5-p} } \Gamma\left( \frac{7-p}2 \right)^\frac{2}{5-p} \lambda_{p+1}^\frac{p-3}{5-p} N^2 V_q T^{p-1}$ \\ [2mm]
\hline
 $\lambda_1$     & $\frac{1}{5-p} { 2^\frac{2(6-p)}{5-p} \pi^\frac{3-p}{5-p} \over (7-p)^\frac{9-p}{5-p} } \Gamma\left( \frac{7-p}2 \right)^\frac{2}{5-p} \lambda_{p+1}^\frac{p-3}{5-p} N^2 V_q T^{p-1}$ \\ [2mm]
\hline
 $\lambda_2$     & $\left( - \frac{2}{5-p} + \frac{2}{7-p} H_\frac{5-p}{7-p} \right) { 2^\frac{2(6-p)}{5-p} \pi^\frac{3-p}{5-p} \over (7-p)^\frac{9-p}{5-p} } \Gamma\left( \frac{7-p}2 \right)^\frac{2}{5-p} \lambda_{p+1}^\frac{p-3}{5-p} N^2 V_q T^{p-1}$ \\ [2mm]
\hline
 $\zeta\tau_\Pi$  & $\left[ \frac{2(p-3)^2 + 2q(5-p)}{(p-q)(5-p)(9-p)} + \frac{2(p-3)^2 + 2q(5-p)}{(p-q)(7-p)(9-p)} H_\frac{5-p}{7-p} \right] { 2^\frac{2(6-p)}{5-p} \pi^\frac{3-p}{5-p} \over (7-p)^\frac{9-p}{5-p} } \Gamma\left( \frac{7-p}2 \right)^\frac{2}{5-p} \lambda_{p+1}^\frac{p-3}{5-p} N^2 V_q T^{p-1}$ \\ [2mm]
\hline
 $\xi_1$     & $\frac{(p-3)^2 + q(5-p)}{(p-q)(5-p)(9-p)} { 2^\frac{2(6-p)}{5-p} \pi^\frac{3-p}{5-p} \over (7-p)^\frac{9-p}{5-p} } \Gamma\left( \frac{7-p}2 \right)^\frac{2}{5-p} \lambda_{p+1}^\frac{p-3}{5-p} N^2 V_q T^{p-1}$ \\ [2mm]
\hline
 $\xi_2$     & $\left[ \frac{[2(p-3)^2 + 2q(5-p)] [(3p^2 - 17p + 18) + 3q(5-p)]}{(p-q)^2 (5-p)(9-p)^2} + \frac{(5-p) [2(p-3)^2 + 2q(5-p)]}{(p-q) (7-p) (9-p)^2} H_\frac{5-p}{7-p} \right]  $  \\ [2mm]
 & $\times { 2^\frac{2(6-p)}{5-p} \pi^\frac{3-p}{5-p} \over (7-p)^\frac{9-p}{5-p} } \Gamma\left( \frac{7-p}2 \right)^\frac{2}{5-p} \lambda_{p+1}^\frac{p-3}{5-p} N^2 V_q T^{p-1}$ \\ [2mm]
\hline
\end{tabular}
\caption{\label{tab: reformulate the results in field theory language} The dual field theory reformulation of the results of compactified Dp-brane. Here $V_q$ is the volume of the compact dimensions of Dp-brane, $\lambda_{p+1} = g_{p+1}^2 N T^{p-3}$ with $g_{p+1}^2 = (2\pi)^{p-2} g_s l_s^{p-3}$. $\lambda_3$ and $\xi_3$ for the compactified Dp-brane are both zero.}
\end{table}

The compactification on the gravity side in holographic relativistic hydrodynamics introduces the volume of the compact directions $V_q$ to balance the dimension of the physical results. $V_q$ in the results shown in \cref{tab: reformulate the results in field theory language} will neutralize the extra mass dimension from $T$ to make the dimension of thermal quantities and transport coefficients to be right.

\section{Summary and discussions} \label{sec: summary}

In this paper, we derive all the second order dynamical transport coefficients for Dp-brane with $q$ directions of their world-volume compactified. To be more specific, the situations considered in this paper include the D4-brane with 2 and 3 directions of its world-volume compactified, D3-brane with 1 and 2 directions compactified, as well as D2-brane with 1 direction compactified. This work can be seen as a generalization of \cite{Wu1807} which considers only the Dp-brane, thus by setting $q=0$ one can reproduce all the results of \cite{Wu1807}. Through \cite{Wu1508,Wu1604,Wu1807} and this work, we have finished the calculation for all the dynamical second order transport coefficients for Dp-brane with or without compactified dimensions.

Through the calculation we can see that the compactification on the world-volume of Dp-brane only affects the scalar perturbation $h$, which can be seen as the spatial trace part of the perturbative metric and relates to the number of the spatial dimensions of the metric. At the first order, $h^{(1)}$ contributes to the spatial diagonal components of the Brown-York tensor on the gravity side and determines the value of $\zeta$. From the boundary fluid viewpoint, $h^{(1)}$ is dual to the viscous term $\partial u$ of the stress tensor. Thus the bulk viscosity $\zeta$ is dependent on the number of compact dimensions $q$. At the second order, the bulk flow viscous terms of $\eta\tau_\pi^*,\ \zeta\tau_\Pi,\ \xi_1,\ \xi_2$ relate with $\partial u$. Their results all contain the numerical factor of $\zeta$ and depend on the number of compact directions $q$. Whereas results of $\eta\tau_\pi,\ \lambda_1,\ \lambda_2$ do not contain $q$ since their viscous tensors do not need to sum the spatial components and are not affected by compactification.

The second order transport coefficients for the compactified Dp-brane also satisfy the Haack-Yarom, Romatschke and Kleinert-Probst relations as in \cite{Wu1807}. Calculation shows the dispersion relations of the D(p-q)-brane do not change compared with Dp-brane case. Although the calculation of the dispersion relations connect with the linear viscous terms $\partial u$ and $D(\partial u)$. From the discussions above, the dispersion relations should depend on the number of compact directions $q$. But the $q$-dependent factors cancel each other in the calculation process making the final results of the dispersion relations contain no $q$.

We know from \cite{Wu1807} that near-extremal black D3-brane will lead to the asymptotically AdS$_5$ black hole after integrating out the unit 5-sphere. Keep on compactifying one or more world-volume directions of the near-extremal black D3-brane equals to compactify the AdS$_5$ black hole. From the results of the compactified D3-brane in this work, one can see that compactification on AdS$_5$ black hole can lead to nonconformal results for the dual fluid's transport coefficients. This reminds us that we can also get nonconformal transport coefficients from compactified AdS black holes of other dimensions. This calculation may give a direct check on the method of obtaining nonconformal hydrodynamical stress-energy tensor from a conformal one that is proposed in \cite{Kanitscheider0901}.

\section*{Acknowledgement}

C. Wu would like to thank Yu Lu for discussions. This work is supported by the Young Scientists Fund of the National Natural Science Foundation of China (Grant No. 11805002).

\appendix

\section{Christoffel symbol and Ricci quantities of the reduction ansatz}

The reduction ansatz used in this work is
\begin{align}
  ds^2 = e^{2\alpha_1 A} g_{MN} dx^M dx^N + e^{2\alpha_2 A}\left( e^{2\beta_1 B} \delta_{mn}dy^m dy^n + e^{2\beta_2 B} L_p^2 d\Omega_{8-p}^2 \right).
\end{align}
We separately denote $\widetilde \Gamma^{\hat M}_{\hat N \hat P}$ and $\Gamma^M_{NP}$ as the Christoffel symbols in 10 dimension and $p-q+2$ dimensional reduced theory. $^{\Omega}\Gamma^a_{bc}$ is the Christoffel symbol on $8-p$ dimensional unit sphere. The Christoffel symbols of the reduction ansatz can be listed as
\begin{align}
  \widetilde\Gamma^M_{NP} &= \Gamma^M_{NP} + \alpha_1 ( \delta^M_N\partial_P A + \delta^M_P\partial_N A - g_{NP}\nabla^M A ), \cr
  \widetilde\Gamma^M_{mn} &= - (\alpha_2\nabla^M A + \beta_1\nabla^M B) e^{(-2\alpha_1 + 2\alpha_2)A + 2\beta_1B}\delta_{mn}, \cr
  \widetilde\Gamma^n_{Mm} &= (\alpha_2\partial_M A + \beta_1\partial_M B) \delta_m^n, \cr
  \widetilde\Gamma^M_{ab} &= - (\alpha_2\nabla^M A + \beta_2\nabla^M B) e^{(-2\alpha_1 + 2\alpha_2)A + 2\beta_2B} \gamma_{ab} L_p^2, \cr
  \widetilde\Gamma^a_{Mb} &= (\alpha_2\partial_M A + \beta_2\partial_M B) \delta^a_b, \cr
  \widetilde\Gamma^a_{bc} &= {}^{\Omega}\Gamma^a_{bc}.
\end{align}
The following relations are useful in the calculation:
\begin{align}
  \widetilde\Gamma^{N}_{MN} &= \Gamma^N_{MN} + (p-q+2) \alpha_1\partial_M A, \cr
  \widetilde\Gamma^{\hat N}_{M\hat N} &= \Gamma^N_{MN} + [(p-q+2)\alpha_1 + (8-p+q)\alpha_2] \partial_M A + [q \beta_1 + (8-p)\beta_2] \partial_M B. \quad
\end{align}
Here $\widetilde\Gamma^{\hat N}_{M\hat N} = \widetilde\Gamma^N_{MN} + \widetilde \Gamma^n_{Mn} + \widetilde \Gamma^a_{Ma}$. The 10 dimensional Ricci tensors are
\begin{align}
  \mathcal R_{MN} =&\; R_{MN} - [(p-q)\alpha_1 + (8-p+q)\alpha_2] \nabla_M \nabla_N A - \alpha_1 g_{MN} \nabla_P \nabla^P A  \cr
  &  - [q \beta_1 + (8-p)\beta_2] \nabla_M\nabla_N B \cr
  &+ \left[ (p-q) \alpha_1^2 + 2(8-p+q) \alpha_1\alpha_2 - (8-p+q) \alpha_2^2 \right] \partial_M A\partial_N A \cr
  & - \left[ (p-q) \alpha_1^2 + (8-p+q) \alpha_1\alpha_2 \right] g_{MN} (\partial A)^2 \cr
  &+ (\alpha_1 - \alpha_2) [ q \beta_1 + (8-p) \beta_2 ] (\partial_M A \partial_N B + \partial_N A \partial_M B) \cr
  &- \alpha_1 [ q \beta_1 + (8-p) \beta_2 ] g_{MN} \partial_P A \partial^P B - \left[q \beta_1^2 + (8-p) \beta_2^2 \right] \partial_M B \partial_N B; \\
  \mathcal R_{mn} =& - \left[ \alpha_2 \nabla^2 A + \beta_1 \nabla^2 B + \left( (p-q) \alpha_1 \alpha_2 + (8-p+q) \alpha_2^2 \right) (\partial A)^2 \right. \cr
  & + \left. \left( (p-q) \alpha_1 \beta_1 + (8-p+2q) \alpha_2 \beta_1 + (8-p) \alpha_2 \beta_2 \right) \partial A \partial B \right. \cr
  & \left.+ \left( q \beta_1^2 + (8-p) \beta_1\beta_2 \right) (\partial B)^2 \right] e^{(-2\alpha_1 + 2\alpha_2) A + 2 \beta_1 B} \delta_{mn}; \\
  \mathcal R_{ab} =&\ (7-p) \gamma_{ab} - \left[ \alpha_2\nabla^2 A + \beta_2 \nabla^2 B + \left( (p-q) \alpha_1 \alpha_2 + (8-p+q) \alpha_2^2\right) (\partial A)^2  \right. \cr
  &\left. + \left( (p-q) \alpha_1\beta_2 + q \alpha_2\beta_1 + (16-2p+q) \alpha_2 \beta_2 \right) \partial A \partial B \right. \cr
  &\left. + \left( q \beta_1 \beta_2 + (8-p) \beta_2^2\right) (\partial B)^2 \right] e^{(- 2 \alpha_1 + 2 \alpha_2) A + 2 \beta_2 B} \gamma_{ab} L_p^2.
\end{align}
From the above we can get the Ricci scalar as
\begin{align}
   \mathcal R =&\; \frac{(7-p)(8-p)}{L_p^2} e^{-2\alpha_2 A - 2\beta_2 B} + e^{-2\alpha_1 A} \left[ R - 2 ((p-q+1) \alpha_1 + (8-p+q) \alpha_2) \nabla^2 A \right. \cr
  &\left. -2 (q \beta_1 + (8-p) \beta_2) \nabla^2 B - \left( (p-q)(p-q+1) \alpha_1^2 + 2(p-q)(8-p+q) \alpha_1\alpha_2 \right.\right.\cr
  &\left.\left. + (8-p+q)(9-p+q) \alpha_2^2 \right) (\partial A)^2 - 2 \left( q(q-p) \alpha_1\beta_1 + (8-p)(p-q) \alpha_1\beta_2 \right.\right. \cr
  &\left.\left. + q(9-p+q) \alpha_2\beta_1 + (8-p)(9-p+q) \alpha_2\beta_2 \right) \partial A \partial B \right. \cr
  &\left. - \left( q(q+1) \beta_1^2 + 2q(8-p) \beta_1 \beta_2 + (8-p)(9-p) \beta_2^2 \right) (\partial B)^2 \right].
\end{align}


\providecommand{\href}[2]{#2}\begingroup\raggedright\endgroup

\end{document}